\documentclass[
 reprint,
superscriptaddress,
 amsmath,amssymb,
 aps,
prl,
]{revtex4-2}

\usepackage{amsmath}
\usepackage{graphicx} 
\usepackage{dcolumn}
\usepackage{bm}
\usepackage{siunitx}
\usepackage{braket}
\begin{document}

\preprint{APS/123-QED}

\title{Entropy of a double quantum dot}

\author{David Kealhofer}
 \email{david.kealhofer@gmail.com}
 \affiliation{%
 Laboratory for Solid State Physics, ETH Z\"urich, CH-8093 Z\"urich, Switzerland
}%
\author{Christoph Adam}
\affiliation{%
 Laboratory for Solid State Physics, ETH Z\"urich, CH-8093 Z\"urich, Switzerland
}%
\author{Max J.\@ Ruckriegel}
\affiliation{%
 Laboratory for Solid State Physics, ETH Z\"urich, CH-8093 Z\"urich, Switzerland
}%
\author{Petar Tomi\'{c}}
\affiliation{%
 Laboratory for Solid State Physics, ETH Z\"urich, CH-8093 Z\"urich, Switzerland
}%
\author{Benedikt Kratochwil}
\affiliation{%
 Laboratory for Solid State Physics, ETH Z\"urich, CH-8093 Z\"urich, Switzerland
}%
\author{Christian Reichl}
\affiliation{%
 Laboratory for Solid State Physics, ETH Z\"urich, CH-8093 Z\"urich, Switzerland
}%
\author{Yigal Meir}
\affiliation{Department of Physics, Ben-Gurion University of the Negev, Beer-Sheva, 84105, Israel}
\author{Werner Wegscheider}
\affiliation{%
 Laboratory for Solid State Physics, ETH Z\"urich, CH-8093 Z\"urich, Switzerland
}%
\affiliation{Quantum Center, ETH Z\"urich, CH-8093 Z\"urich, Switzerland}
\author{Thomas Ihn}
\affiliation{%
 Laboratory for Solid State Physics, ETH Z\"urich, CH-8093 Z\"urich, Switzerland
}%
\affiliation{Quantum Center, ETH Z\"urich, CH-8093 Z\"urich, Switzerland}
\author{Klaus Ensslin}
\affiliation{%
 Laboratory for Solid State Physics, ETH Z\"urich, CH-8093 Z\"urich, Switzerland
}%
\affiliation{Quantum Center, ETH Z\"urich, CH-8093 Z\"urich, Switzerland}

\date{\today}

\begin{abstract}
We use charge sensing to detect entropy changes in a double quantum dot defined by electrostatic gating of a GaAs/AlGaAs heterostructure.
This system can be tuned to be two separate systems, like two independent, artificial atoms, or a single coherent system, like a molecule.
We study entropy changes in both regimes due to changes in the occupation of the system.
First we recover the single-dot result for each dot, that the occupation of the dot by a single electron corresponds to an increase in the entropy of $k_{\mathrm{B}} \log 2$.
Next we examine two different charge transitions in the ``molecular'' regime, and how it reveals itself in terms of the measured entropy.
We also uncover a realization of Pauli blockade that clutters the entropy signal.
By applying a rate equation model, we demonstrate the effect's nonequilibrium origins and exclude it from the analysis of the system's entropy.
Understanding these experiments in this simplest coupled system enables the study of the entropy in other, more complicated coupled quantum systems, such as ones with topological or highly entangled ground states.
\end{abstract}

\maketitle

Entropy is a fundamental thermodynamic property of physical systems.
An extensive quantity, proportional to the system's volume, it can be prohibitively difficult to measure in micrometer- or nanometer-scale systems using the methods successful in macroscopic ones, such as by using heat capacity.
A recently invented technique, however, measures the entropy due to an electron's occupation of a single quantum dot, using a Maxwell relation to transpose the measurement into a charge sensing experiment \cite{hartman_direct_2018,child_robust_2022}.
This technique has been extended to more complicated physical problems, i.e.\@ to study the entropy of a quantum dot strongly coupled to a lead \cite{child_entropy_2022}, as well as to quantum dots in other materials systems \cite{christoph_2025}.

This technique's promise lies in its application to quantum systems coupled to such single quantum dots.
It has been proposed that a quantum dot coupled to a hybrid semiconductor--superconductor system can detect the fractional entropy of a Majorana zero mode and, importantly, discriminate between the topologically nontrivial Majorana zero mode and topologically trivial Andreev bound state \cite{sela_detecting_2019}, necessary for using such a system for topological quantum computation \cite{nayak_nonabelian_2008, lutchyn_majorana_2018}.
Similarly, the natural extension of this technique to quantum dot arrays should provide a powerful tool to study simulated quantum systems.

Here we study the entropy of the simplest coupled quantum dot system: the double quantum dot (DQD).
This system has the advantage of simplicity and straightforward comparison to the previously studied single dot.
The DQD, nevertheless, can be tuned from two non-interacting artificial atoms into a single artificial molecule, which is to say it is a convenient model system that allows us to extend our technique from single- to multiple-particle physics.

\begin{center}
\begin{figure}[tb]
    \includegraphics[scale= 1]{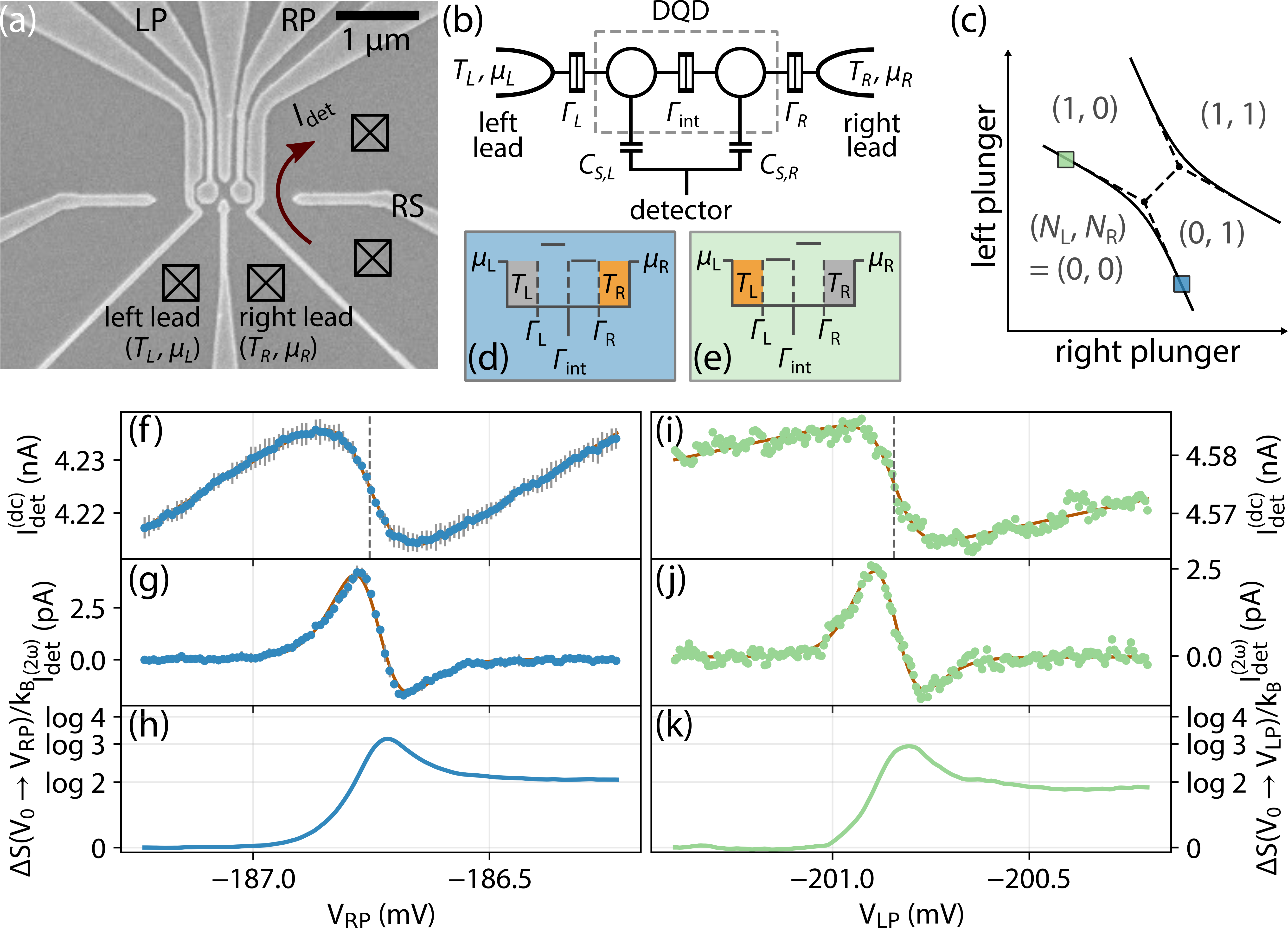}
    \caption{\label{fig:1} (a) Scanning electron micrograph of a nominally identical device. 
    $V_{\mathrm{LP}}$ and $V_{\mathrm{RP}}$, applied with respect to the 2DEG, tune the DQD occupation, which is measured by $I_{\mathrm{det}}$ through the nearby QPC.
    (b) Device schematic, including the interdot coupling strength $\Gamma_{\mathrm{int}}$ and the coupling to the left and right leads, $\Gamma_\mathrm{L}$ and $\Gamma_{\mathrm{R}}$.
    The leads are thermal baths of electrons with temperatures $T_\mathrm{L}$ and $T_\mathrm{R}$ and chemical potentials $\mu_{\mathrm{L}}$ and $\mu_{\mathrm{R}}$. 
    The capacitive coupling of the DQD to the detector differs for the left ($C_{\mathrm{S,L}}$) and right dots ($C_{\mathrm{S,R}}$).
    (c) Charge stability diagram schematic.
    Transitions between regions of stable charge occur along the dashed ($\Gamma_\mathrm{int} = 0$) or solid ($\Gamma_\mathrm{int} > 0$) lines.
    The difference is greatest near the triple points (solid circles).
    (d) Level diagram for the $(N_\mathrm{L}, N_{\mathrm{R}}) = (0,0) \rightarrow (0,1)$ transition with the left dot far detuned [blue box in (c)].
    The only relevant coupling is $\Gamma_\mathrm{R}$, the only relevant temperature is $T_\mathrm{R}$.
    (e) Level diagram for the $(0,0) \rightarrow (1,0)$ transition with the right dot far detuned [green box in (c)].
    (f) The dc detector current in configuration (d), using $V_\mathrm{RP}$ to sweep across the transition.
    The points are averaged from multiple sweeps, the error bars are the standard deviation, and the line is a fit to Eq.~(\ref{eqn:dc_fit}).
    A dashed line marks $V_\mathrm{mid}$.
    (g) The detector current measured at the second harmonic of the heating current, configuration (d). The points are data. The line is a fit to Eq.~(\ref{eqn:octave_fit}).
    (h) Cumulative integral of $\Delta S$ with respect to $V_{\mathrm{RP}}$. [See Eq.\@~(\ref{eqn:integrate}).]
    (i-k) The same as (f-h), instead using $V_{\mathrm{LP}}$ to traverse the 
    $(0,0) \rightarrow (1,0)$ transition, as in (e).}
\end{figure}
\end{center}

Our DQD, in a dilution refrigerator with a mixing chamber temperature of $\SI{8}{mK}$, is formed by using nanoscopic gates to deplete a two-dimensional electron gas (2DEG) at a GaAs/AlGaAs heterointerface \SI{110}{nm} below the wafer's surface \cite{supp}.
The gate layout is shown in Fig.~\ref{fig:1}(a).
The right and left leads, regions of the 2DEG connected to ohmic contacts, act as thermal reservoirs of electrons, with well-defined temperatures and chemical potentials, as depicted in Fig.~\ref{fig:1}(b).
Each reservoir is connected to two ohmic contacts; ac current through each pair controls the temperature of each reservoir.
The DQD occupation is detected by measuring the dc current through a quantum point contact (QPC) capacitively coupled to the DQD, whose conductance is tuned between full pinchoff and the first plateau \cite{field_measurements_1993}.

Starting with the DQD unoccupied, the system's average occupation can be increased one electron at a time by tuning the plunger voltages $V_{\mathrm{LP}}$ and $V_{\mathrm{RP}}$ to bring an energy level of the DQD down to the Fermi energy of the leads, as depicted in the idealized charge stability diagram of Fig.~\ref{fig:1}(c).
When the energies of the dots are detuned from each other, we access single-dot-like transitions and can study the change in the Boltzmann entropy $\Delta S$ in those configurations, as depicted for the $(N_L,N_R) = (0,0) \rightarrow (0,1)$ [Fig.~\ref{fig:1}(d)] and $(0,0) \rightarrow (1,0)$ [Fig.~\ref{fig:1}(e)] transitions.
With this large detuning, the left--right charge basis is natural
and only one tunneling rate relevant, e.g.\@ $\Gamma_\mathrm{R}$ in Fig.~\ref{fig:1}(d). 
Then, if the transition is thermally broadened (i.e., for a right dot transition, [Fig.~\ref{fig:1}(d)]\@~$\Gamma_\mathrm{R} \ll k_{\mathrm{B}} T / \hbar$), we can analyze $\Delta S$ as for a single dot as follows.

There are two related strategies for determining $\Delta S$~\cite{hartman_direct_2018, child_entropy_2022, child_robust_2022}.
One is based on fitting the second harmonic of the heating current~\cite{endmatter}.
Consider a transition on the right dot [Fig.~\ref{fig:1}(d)], caused by sweeping $V_{\mathrm{RP}}$.
With an ac current heating the right lead at a frequency $\omega$, the temperature of the lead oscillates at the frequency $2\omega$, with an amplitude $\delta T$.
The Boltzmann entropy is encoded in $V_\mathrm{mid}$, where it is equally likely for the electron to occupy the lead as the dot, marked with a dashed vertical line in the dc charge detection signal shown in Fig.~\ref{fig:1}(f).
This depends on the degeneracy of each state \cite{gustavsson_electron_2009, hofmann_measuring_2016},
and its variation with temperature is precisely what gives the entropy change between the two charge states, i.e.\@~$\partial V_{\mathrm{mid}}/\partial \theta =  -\Delta S / k_{\mathrm{B}}$.

We can access the derivative $\partial V_{\mathrm{mid}}/\partial \theta$, and so measure $\Delta S$, by measuring the current through the detector at $2\omega$, 
$I^{(2\omega)}_{\mathrm{det}} = (\partial I^{(\mathrm{dc})}_{\mathrm{det}}/ \partial T) \delta T$ \cite{hartman_direct_2018}, or, explicitly,
\begin{align}
 I^{(2 \omega)}_{\mathrm{det}} = I^{(2\omega)}_0 \!\left( \frac{V_{\mathrm{RP}} - V_{\mathrm{mid}}}{2 \theta_{\mathrm{RR}}} - \frac{\Delta S}{2 k_{\mathrm{B}}}\right)\! \cosh^{-2} \frac{V_{\mathrm{RP}} - V_{\mathrm{mid}}}{2 \theta_{\mathrm{RR}}}. \label{eqn:octave_fit}
\end{align}
Important is $\theta_{\mathrm{RR}}$, the temperature in units of $V_{\mathrm{RP}}$, i.e.\@~$k_{\mathrm{B}} T / e \alpha_{\mathrm{RR}}$, where $\alpha_{\mathrm{RR}}$ is the lever arm for $V_{\mathrm{RP}}$ on the right dot and $k_{\mathrm{B}}$ is the Boltzmann constant.
Figure~\ref{fig:1}(g) shows data collected simultaneously with that in Fig.~\ref{fig:1}(f), averaged and fit with Eq.~(\ref{eqn:octave_fit}); we fit $\theta_{\mathrm{RR}} = (36.03 \pm 0.18)~\mathrm{\mu}$V, corresponding to a heated reservoir temperature of $T_\mathrm{R} = (55.61 \pm 0.03)$ mK, and $\Delta S/k_{\mathrm{B}} = 0.726 \pm 0.012$, i.e.\@ within $10\%$ of the expected change in Boltzmann entropy, $\Delta S/k_\mathrm{B} = \log \Omega_{(0,1)}/\Omega_{(0,0)} =\log 2/1 \approx 0.693$, where $\Omega$ is the number of available microstates; since the singly occupied charge state is spin-degenerate.

The other way to find $\Delta S$ is to integrate another Maxwell relation:
\begin{align}
    \Delta S_{\epsilon_1 \rightarrow \epsilon_2}/k_{\mathrm{B}} &= \int_{\epsilon_1}^{\epsilon_2} \mathrm d \mu \, \frac{\partial N_R}{\partial T_R} = \frac{1}{2}\int_{V_{\mathrm{RP},1}}^{V_{\mathrm{RP},2}} \mathrm d V_{\mathrm{RP}} \, \frac{I^{(2 \omega)}_{\mathrm{det}}}{\theta_{\mathrm{RR}} I^{(2 \omega)}_0}, \label{eqn:integrate}
\end{align}
where the second equality rewrites the integral in experimental terms.
In the weak coupling regime, the integrated $\Delta S$ reaches an intermediate peak of $\Delta S = k_{\mathrm{B}} \log 3$, which measures the system's total degeneracy when the $(0,0)$ state (degeneracy of 1) is resonant with the $(0,1)$ state (degeneracy of 2)~\cite{child_entropy_2022}.
We treat the presence of this peak as a signature of the weak coupling regime, that is, an indicator of the applicability of the fitting procedure.
Figure~\ref{fig:1}(h) shows the cumulative integral of the data plotted in Fig.~\ref{fig:1}(g), using the values of $\theta_{\mathrm{RR}}$ and $I^{(2\omega)}_0$ from the fit.
$\Delta S$ reaches a final value of $ 0.72  k_{\mathrm{B}}$. 

The same experiment on the left dot [see Figs.~\ref{fig:1}(c,e)] produces the same results.
Figures~\ref{fig:1}(i-k) show the corresponding dc, second harmonic, and integrated traces [compare Figs.~\ref{fig:1}(f-h)].
In this case, we fit $\theta_{\mathrm{LL}} = (39.9 \pm 0.7)~\mathrm{\mu}$V, or $T_\mathrm{L} = (52.8 \pm 0.9)$ mK, and $\Delta S/k_{\mathrm{B}} = 0.750 \pm 0.042$.
The integrated entropy, $\Delta S/k_{\mathrm{B}} = 0.64$, also agrees with $\log 2$ to within $10\%$. 
In summary, for either dot, whether fitting or integrating, we recover the expected result $\Delta S(N_i = 0 \rightarrow 1) = k_{\mathrm{B}} \log 2$, reflecting the spin degeneracy of the unoccupied level in a single quantum dot.

\begin{center}
\begin{figure*}[tb]
    \includegraphics[scale= 1]{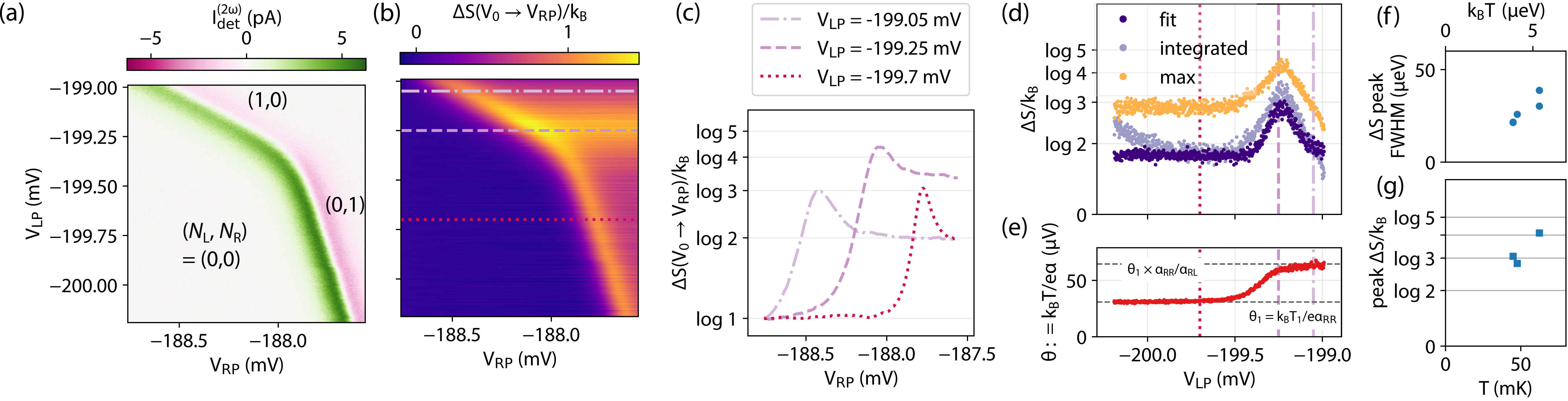}
    \caption{\label{fig:2} Near the triple point, first carrier transition.
    (a) $I^{(2\omega)}_{\mathrm{det}}$ plotted as a function of $V_{\mathrm{RP}}$ and $V_{\mathrm{LP}}$.
    (b) Cumulative integral of $\Delta S$ with respect to $V_{\mathrm{RP}}$, starting from $(0,0)$.
    At every value of $V_{\mathrm{LP}}$, $I_0 \theta$ is found from a fit to $I^{(2\omega)}_{\mathrm{det}}(V_{\mathrm{RP}})$.
    Cuts at constant $V_{\mathrm{LP}}$ are indicated.
    (c) Line cuts from (b) indicated by the dashed lines.
    (d) Fitted $\Delta S$ across the transition for every $V_{\mathrm{LP}}$ in (a).
    The peak value for each integration trace is also plotted, which records an extra degeneracy of one throughout.
    (e) Fitted $\theta = k_{\mathrm{B}} T/ e \alpha$ as a function of $V_{\mathrm{LP}}$.
    Also shown is a dotted line showing an average value $\theta_1$ of $\theta$ at the $(0,0)\rightarrow (0,1)$ transition, far from the triple point, and a dotted line showing $\theta_1 \times \alpha_{\mathrm{RR}}/\alpha_{RL}$.
    The ratio of lever arms is found independently, using the charge stability diagram.
    As in (d), the values of $V_{\mathrm{LP}}$ plotted in (c) are indicated by vertical dashed lines.
    (f) The height and (g) width of the peak in (d) found for different heater powers, plotted against the temperature of the lead, also shown as a thermal energy; these panels share an $x$-axis.
    }
\end{figure*}
\end{center}

To move beyond single-dot behavior, we tune the DQD to a triple point, where three stable charge configurations are energetically close, and charge transfer between the dots can be as important as charge transfer from the leads.
The situation here is more complicated.
For appreciable $\Gamma_{\mathrm{int}}$, the left--right basis becomes less meaningful, which poses a problem for analyzing $I^{(2\omega)}_{\mathrm{det}} = (\partial I^{(\mathrm{dc})}_{\mathrm{det}}/ \partial T) \delta T$ with all the relevant $\mathrm{L,R}$ subscripts.
Moreover, $\Gamma_\mathrm{L}$, $\Gamma_\mathrm{R}$, and $\Gamma_\mathrm{int}$, often difficult to measure directly, are unavoidably tuned by the plunger voltages, affecting $T_i$ and $\delta T_i$.
The simplest approach is nevertheless to modulate the temperature of only one lead, leaving witnesses to possible artifacts.
If the tunneling rate to the heated lead is too low, i.e.\@ the effective rate to the other lead is comparable, $\delta T\rightarrow 0$, $I^{(2\omega)}_{\mathrm{det}}\rightarrow 0$, and $\Delta S$ cannot be extracted~\cite{supp}.
If it is too high, strong lead--DQD coupling results in suppression of the peak in the integrated entropy~\cite{child_entropy_2022}.

Figure~\ref{fig:2}(a) shows $I^{(2\omega)}_{\mathrm{det}}$ near the $(0,0)$, $(1,0)$, $(0,1)$ triple point, with $T_\mathrm{R}$ modulated [compare Fig.~\ref{fig:1}(c)].
For every $V_{\mathrm{LP}}$ we fit $I^{(2\omega)}_{\mathrm{det}}$ as a function of $V_{\mathrm{RP}}$ to Eq.~(\ref{eqn:octave_fit}) and use the fitted values of $\theta$ and $I^{(2\omega)}_{0}$ to integrate $\Delta S$ with respect to $V_{\mathrm{RP}}$, i.e.\@ $\Delta S(V_0 \rightarrow V_{\mathrm{RP}})$.
Three representative traces, indicating transitions near and on either side of the triple point, are marked with dashed lines on Fig.~\ref{fig:2}(b) and shown in Fig.~\ref{fig:2}(c).
Each trace resembles Fig.~\ref{fig:1}(h), with a maximum at the transition, and a final plateau at the total $\Delta S$ between the two charge states.
Farther from the triple point, at $V_{\mathrm{LP}} = -199.05$ mV and $V_{\mathrm{LP}} = -199.7$ mV, $\Delta S$ is $k_{\mathrm{B}} \log 2$, with a  $k_{\mathrm{B}} \log 3$ maximum at the transition.
Near the triple point, both the total $\Delta S$ and its maximum value are higher.

For each $V_{\mathrm{LP}}$ in Fig.~\ref{fig:2}(a), $I^{(2\omega)}_{\mathrm{det}}$ is fit with respect to $V_{\mathrm{RP}}$, and the resulting $\Delta S$ is plotted in Fig.~\ref{fig:2}(d).
Values of $\Delta S$ found from the fit and integration demonstrate a peak near the triple point.
(At the beginning and end of the $V_{\mathrm{LP}}$ range, the signal runs close to an integration limit and causes deviation from the fitted value.)
Throughout, the maximum of the integrated trace follows $d_1+1$, where $d_1$ is the degeneracy of the one-electron state.

The evolution of $\theta$, plotted in Fig.~\ref{fig:2}(e), confirms that tunneling via $\Gamma_\mathrm{R}$ is the only relevant coupling to the system.
When $V_{\mathrm{LP}}$ is very negative, i.e.\@ for the $(0,0) \rightarrow (0,1)$ transition, $\theta = 30.9~\mathrm{\mu}$V, which we call $\theta_1$, corresponding to $T = 47.6$ mK, shown with a dotted line in Fig.~\ref{fig:2}(e). 
For this transition, the relevant lever arm is $\alpha_{\mathrm{RR}}$, i.e.\@ for RP acting on the right dot.
For the $(0,0) \rightarrow (1,0)$ transition, however, we are considering a transition on the left dot.
By working at fixed $V_{\mathrm{LP}}$ and fitting with respect to $V_{\mathrm{RP}}$, the relevant lever arm is $\alpha_{\mathrm{RL}}$, i.e.\@ for RP acting on the left dot.
The ratio $\alpha_\mathrm{RR} / \alpha_\mathrm{RL}$, however, we can determine from the charge stability diagram.
The quantity $\theta_1 \times \alpha_\mathrm{RR} / \alpha_\mathrm{RL}$ is also plotted in Fig.~\ref{fig:2}(e), and it matches the value $\theta$ converges to for the $(0,0) \rightarrow (1,0)$ transition.
This demonstrates that the temperature is constant across the range of $V_{\mathrm{LP}}$ shown and thus excludes that tunneling via $\Gamma_\mathrm{L}$ contributes in some spurious way to $I^{(2\omega)}_{\mathrm{det}}$.

What is the origin of the increased degeneracy?
Figures~\ref{fig:2}(f) and (g) show that the region of higher $\Delta S$, that is, the peak in Fig.~\ref{fig:2}(d), grows in width and height as $T_\mathrm{R}$ is increased.
(For analysis at other temperatures, see \cite{supp}.)
As shown in Fig.~\ref{fig:2}(f), the thermal energy scale is smaller than the width of these features, but comparable to the tunnel coupling 5.8 $\mathrm{\mu}$eV: at $k_{\mathrm{B}} T = 5.4~\mathrm{\mu}$eV, the fully-developed height of the feature, $\Delta S = k_{\mathrm{B}} \log 4$, corresponds to the combined degeneracy of the bonding and antibonding states of the DQD. 
Put differently, the increased degeneracy appears where the DQD level spacing is comparable to the lead's $k_B T$.
When $k_\mathrm{B} T$ is large enough, the measured $\Delta S$ reflects the full degeneracy of the system, with two twofold-degenerate levels, $k_\mathrm{B} \log 4$ [Fig.~\ref{fig:2}(g)].

\begin{center}
\begin{figure}[tb]
    \includegraphics[scale= 1]{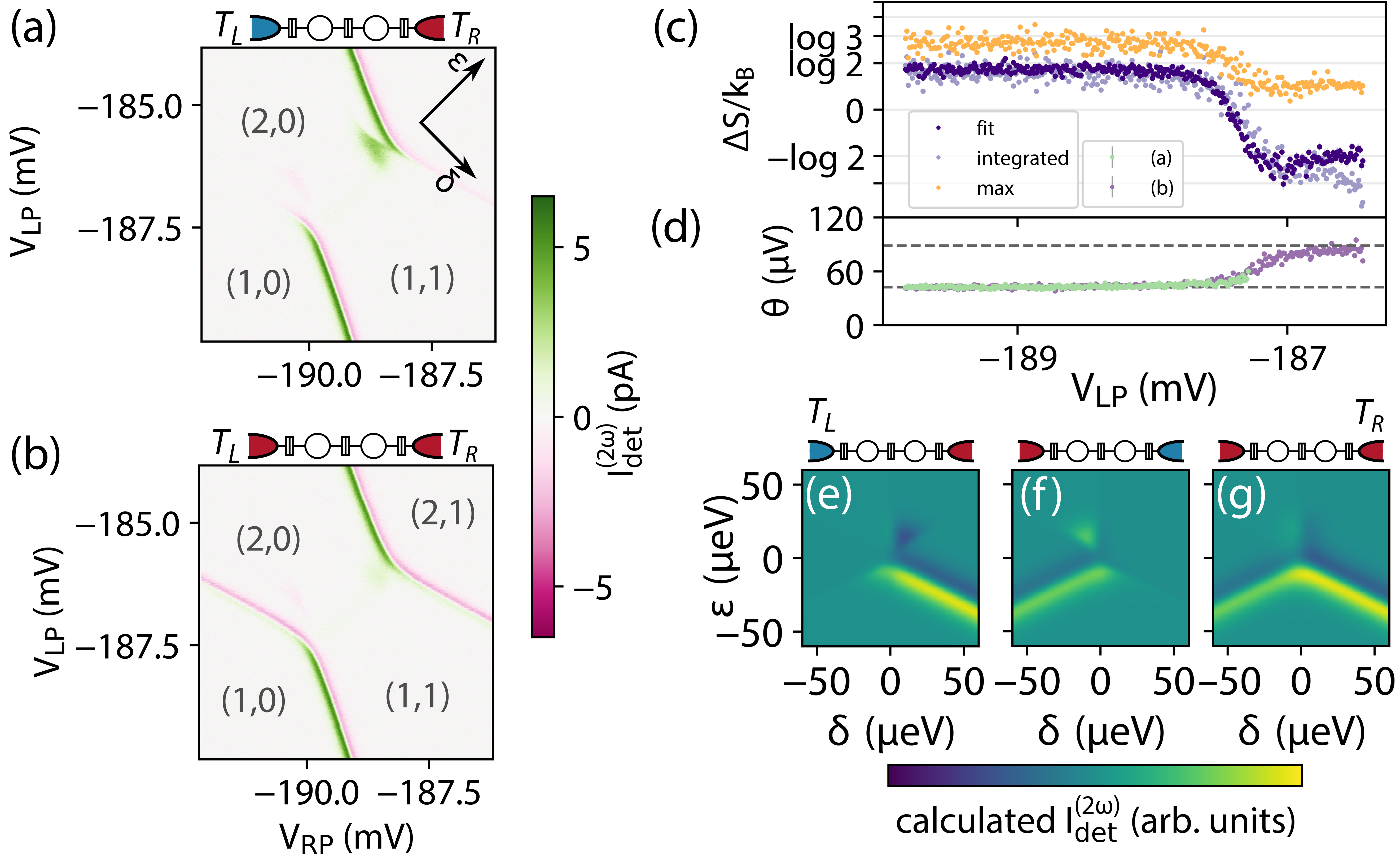}
    \caption{\label{fig:3} Transitions between $(1,0)$, $(1,1)$, $(2,0)$, and $(2,1)$.
    $I^{(2\omega)}_{\mathrm{det}}$ with (a) only the right lead and (b) both leads heated (depicted schematically).
    (c) $\Delta S/k_{\mathrm{B}}$ for transitions out of $(1,0)$ in (b), plotted against $V_{\mathrm{LP}}$.
    The fitted, integrated, and resonant peak (``max'') values agree.
    [Compare Fig.~\ref{fig:2}(d).]
    (d) $\theta := k_{\mathrm{B}} T / e \alpha $ found from fitting Eq.~(\ref{eqn:octave_fit}) to the transitions out of $(1,0)$ in (a) and (b).
    Horizontal lines indicate the constant-temperature criterion far from the triple points.
    (e-g) $I^{(2\omega)}_{\mathrm{det}}$ calculated from a rate equation and simplified DQD model, with (e) right lead, (f) left lead, and (g) both leads heated, plotted against dots' average energy $\epsilon$ and the detuning $\delta$.
    [The directions of increasing $\delta$ and $\epsilon$ are shown relative to the voltage axes in panel (a).]
    }
\end{figure}
\end{center}

While there is no single quantum dot analogue of the above experiment, it is still a single-particle effect. 
To go further, we study the transitions between the states $(1,0)$, $(1,1)$, $(2,0)$, and $(2,1)$.
In Fig.~\ref{fig:3}(a), $I^{(2\omega)}_{\mathrm{det}}$ is plotted for an experiment like that in Fig.~\ref{fig:2}(a), the key difference the charge states.
The right lead is heated, as shown schematically, and transitions of $N_\mathrm{R} = 0 \rightarrow 1$ appear as stripes in $I^{(2\omega)}_{\mathrm{det}}$.

To analyze $\Delta S$ near the triple point, we build on the approach of Fig.~\ref{fig:2}.
There the experiment is controlled by only one lead: $\Gamma_\mathrm{R} \gg \Gamma_\mathrm{L}$ and $\hbar \Gamma_\mathrm{int} \gtrapprox k_{\mathrm{B}} T_R$, as confirmed by the evolution of the ``max'' trace [Fig.~\ref{fig:2}(d)].
Here we examine the right dot--right lead and left dot--left lead transitions in a more symmetric tuning of $\Gamma_\mathrm{R}$ and $\Gamma_\mathrm{L}$.
Figure~\ref{fig:3}(b) shows $I^{(2\omega)}_{\mathrm{det}}$ with both $T_\mathrm{R}$ and $T_\mathrm{L}$ elevated and modulated.
The $N_\mathrm{R} = 0 \rightarrow 1$ transitions are still visible, and so are the $N_\mathrm{L} = 1 \rightarrow 2$ transitions.

As for $\Delta S$, Fig.~\ref{fig:3}(c) shows the results from fitting [Eq.~(\ref{eqn:octave_fit})] and integrating [Eq.~(\ref{eqn:integrate})] $I^{(2\omega)}_{\mathrm{det}}$ in Fig.~\ref{fig:3}(b) for the transitions from $(1,0)$.
Far from the triple points, $\Delta S$ agrees with the single-particle picture.
For the $(1,0)\rightarrow (1,1)$ transition, the fitted $\Delta S/k_{\mathrm{B}} = 0.60 \pm 0.05$, and the integrated $\Delta S/k_{\mathrm{B}} = 0.56 \pm 0.11$, consistent with a change of $k_{\mathrm{B}} \log 2$ . 
(Average is over $V_{\mathrm{LP}} < -188.4$ mV; uncertainty is the standard deviation.)
After passing through the triple point, we fit $\Delta S/k_{\mathrm{B}} = -0.71 \pm 0.074$ for the $(1,0)\rightarrow (2,0)$ transition, where we expect $\Delta S = - k_{\mathrm{B}} \log 2$ as the electron enters a singlet state.
(Average is for $V_{\mathrm{LP}} > -186.8$ mV.)
The evolution of $\theta$ [Fig.~\ref{fig:3}(d)] demonstrates that, in this tuning, the temperature of the left and right leads are identical with and without heating of the left lead: as in Fig.~\ref{fig:2}(e), the dashed lines show the equal temperature criterion away from the triple points.

Absent from Fig.~\ref{fig:2}(a), a triangular feature appears here, of $I^{(2\omega)}_{\mathrm{det}}>0$, near the triple point at $(1,1)$, $(2,0)$, and $(2,1)$ when only the right lead temperature is modulated.
(When the left lead temperature is modulated, a similar triangle appears near the other triple point; see~\cite{supp}.)
When both leads are heated, the triangle disappears.
This persists for all the interdot tunnel couplings measured ~\cite{supp}.
Meanwhile, such triangles never appear at the charge configurations shown in Fig.~\ref{fig:2}(a).

These triangles defy the interpretation of the integral of $I^{(2\omega)}_{\mathrm{det}}$ as the system's entropy, beholden only to changes of the charge configuration: integration paths exist, even changing the chemical potential of only a single dot, that result in differing ostensible $\Delta S$.
For this reason it is important to understand the triangles' origins.

We apply a rate equation to model the behavior of a simulated $I^{(2\omega)}_{\mathrm{det}}$ \cite{supp}.
The calculation reveals that the lead and interdot couplings cannot conspire alone to produce the triangles.
Instead, key ingredients are a slow relaxation rate $\Gamma_\mathrm{r}$ from the excited to the ground state of the system, and the detector's differing sensitivities to charge on the left versus right dot [cf.~Fig.~\ref{fig:1}(b)].
Figures~\ref{fig:3}(e-g) show calculated $I^{(2\omega)}_{\mathrm{det}}$ for configurations like those in Fig.~\ref{fig:3}(a-b), with $\hbar \Gamma_\mathrm{r} = 1~$neV ($\Gamma_\mathrm{r} \approx 1.5$ MHz) and $I^{(\mathrm{dc})}_{\mathrm{R},0} = 1.7 \times I^{(\mathrm{dc})}_{\mathrm{L},0}$, similar to the experiments [compare Fig.~\ref{fig:1}(g) and (j)].
In Fig.~\ref{fig:3}(e), with the right lead heated, a triangular feature appears, but of the wrong sign and on the wrong side of the $\delta = 0$ line---we do not observe this in the experiments.
Figure~\ref{fig:3}(f) depicts $I^{(2\omega)}_{\mathrm{det}}$ with the left lead heated.
Here, the triangle has the correct sign, appears at the correct detuning, and, like in the data, has one side sharp, one side thermally smeared.
Finally, Fig.~\ref{fig:3}(g) shows a calculation with both leads heated.
No triangles appear, instead only a yellow-and-blue stripe along $\delta = 0$.

The triangles are a manifestation of Pauli blockade.
They occur by a process that drives the electron into the right dot via the DQD's excited state, generating nonzero $\mathrm d N_\mathrm{R} / \mathrm d T_\mathrm{R(L)}$~\cite{supp}.
In the $(2,0)$ state, relaxation from the excited states---the degenerate singlet and triplet $(1,1)$ states---into the ground state $(2,0)$ singlet is slow;
on the $(1,1)$ side of $\delta = 0$, the excited state is the $(2,0)$ singlet, relaxation to the degenerate $(1,1)$ ground state is fast.
The triangles appear only on the $(2,0)$ side of the $\delta = 0$ line, not on the $(1,1)$ side, and not near the triple point at $(0,0)$: only with suppression of the relaxation by a spin selection rule is the relaxation rate slow enough to observe them.
In other words, the triangle is a nonequilibrium phenomenon, beyond the equilibrium thermodynamics that give the Maxwell relation that connects $I^{(2\omega)}_{\mathrm{det}}$ and $\Delta S$.

In conclusion, we have measured $\Delta S$ in the single-carrier ``atomic'' and ``molecular'' configurations of a DQD, as well as across the transitions in the vicinity of $(1,0),(1,1),(2,0),(2,1)$, where we disentangle nonequilibrium effects---the relaxation of the excited state---from the desired entropy signal.
These results can guide future work to understand the ground states of more complicated coupled quantum systems.

The authors acknowledge financial support by the H2020 European Research Council (ERC) Synergy Grant under grant agreement 951541 and the Swiss National Science Foundation, grant 200021\_204968.
\bibliography{dqd_entropy_bib}

\appendix*
\section*{End Matter}
\textbf{From dc charge detection to measuring entropy.} Measurements in this work rely on fitting the dc component of the current through the QPC detector, $I^{(\mathrm{dc})}_{\mathrm{det}}$.
Consider a transition on the right dot [Fig.~\ref{fig:1}(d)], caused by sweeping the right plunger voltage, $V_{\mathrm{RP}}$.
When the coupling to the lead is weak, i.e.\@ $\Gamma_\mathrm{R} \ll k_{\mathrm{B}} T_\mathrm{R}$, transferring an electron between the right lead and the right dot [Fig.~\ref{fig:1}(d)] causes a step in the detector current,
\begin{align}
I^{(\mathrm{dc})}_{\mathrm{det}} = -I^{(\mathrm{dc})}_{\mathrm{R},0} \tanh \frac{V_{\mathrm{RP}} - V_{\mathrm{mid}}}{2 \theta_{\mathrm{RR}}} + \gamma V_{\mathrm{RP}} + I_0, \label{eqn:dc_fit}
\end{align}
where $I_0$ is an offset, $\gamma$ accounts for the capacitive coupling of $V_{\mathrm{RP}}$ to the detector, and $I^{(\mathrm{dc})}_{\mathrm{R},0}$ parameterizes the detector's sensitivity to a change in occupation on the right dot.
The $\tanh$ term describes the total change in the quantum dot's charge, connecting the thermodynamic quantities $N$, $T$, and $\mu$ to the detector current.
Figure~\ref{fig:1}(f) shows data averaged from multiple sweeps of $V_{\mathrm{RP}}$ in a configuration like Fig.~\ref{fig:1}(d) and a fit to Eq.~(\ref{eqn:dc_fit}); the same experiment is shown for a left dot--left lead transition in Fig.~\ref{fig:1}(g).
From data like these, we can fit, using Eq.~\ref{eqn:dc_fit}, the relevant parameters $I^{(\mathrm{dc})}_{i,0}$, $V_{\mathrm{mid}}$, and $\theta_{ij}$ (where $i$ and $j$ are R or L, depending on the transition and which plunger gate is used), as well as the parameters not directly relevant to the analysis of the entropy, the cross-capacitance $\gamma$ and the setpoint current $I_0$.

At the same time, we measure the current through the QPC detector at twice the frequency of the heating excitation in the leads, which we call $I^{(2\omega)}_{\mathrm{det}}$.
Using Eq.~\ref{eqn:dc_fit} to evaluate the derivative $\partial V_{\mathrm{mid}}/\partial \theta$ gives Eq.~\ref{eqn:octave_fit}, which we use throughout to fit $I^{(2\omega)}_\mathrm{det}$.
In so doing, the overall ``sensitivity'' $I^{(2\omega)}_0$, the entropy change $\Delta S$, the temperature in units of voltage $\theta_{\mathrm{ij}}$, and the midpoint voltage $V_{\mathrm{mid}}$ are fit parameters; the lever arm, needed to convert $\theta_{ij}$ to $T_{i(j)}$, is derived from a measurement at finite bias~\cite{supp}.
(Note that $I^{(2\omega)}_0$ is as deserving of a dot subscript as $I^{(\mathrm{dc})}_{i,0}$---$I^{(2\omega)}_0$ depends on the temperature excitation $\delta T$ and the dc sensitivity of the detector to a transition on either dot (schematically depicted in Fig.~\ref{fig:1}(b) as $C_{\mathrm{S,L(R)}}$).
But when we discuss the derivative, $I^{(2\omega)}_{\mathrm{det}} = (\partial I^{(\mathrm{dc})}_{\mathrm{det}}/ \partial T) \delta T$, we suppress the subscript to better discuss the continuous evolution of $I^{(2\omega)}_0$ through the triple point.)
Finally, we monitor, as discussed in the Supplemental Material, the ``first harmonic,'' i.e.\@ the ac current through the detector at the heater modulation frequency, $I_{\mathrm{det}}^{(\mathrm{ac})}$~\cite{supp}.

%%% start SM
\newpage

\renewcommand{\thesection}{S\arabic{section}}  
\renewcommand{\thetable}{S\arabic{table}}  
\renewcommand{\thefigure}{S\arabic{figure}}
\renewcommand{\theequation}{S\arabic{equation}}
\renewcommand{\thepage}{S\arabic{page}}

\setcounter{page}{1}
\setcounter{figure}{0}

\onecolumngrid

\section*{Supplemental Material}

\section{\label{sec:device_fabrication} Device fabrication}
The device is fabricated from a GaAs/AlGaAs heterostructure using photolithography and electron beam lithography.
A dilute piranha etch defines the mesa; a stack of (top to bottom) Au/Ni/Au/Ge, deposited by electron beam evaporation and then annealed in forming gas, is used for ohmic contact; and the nanoscopic Schottky gates of the quantum dots, sensor quantum point contacts (QPCs), and heater QPCs are Au/Ti, also deposited by electron beam evaporation.
An additional fanout layer of Au/Ti connects the gates and contacts to bondpads, from which the device is wirebonded to a printed circuit board for mounting in the dilution refrigerator.

\section{Overview of the technique}
These experiments were performed in a dilution refrigerator with a mixing chamber temperature of 8 mK.
Besides commercial lock-in amplifiers and dc multimeters, we use homemade dc voltage sources and current--voltage converters.

\subsection{Description of the device}
\begin{center}
\begin{figure}[b!]
    \includegraphics[width= 0.5\textwidth]{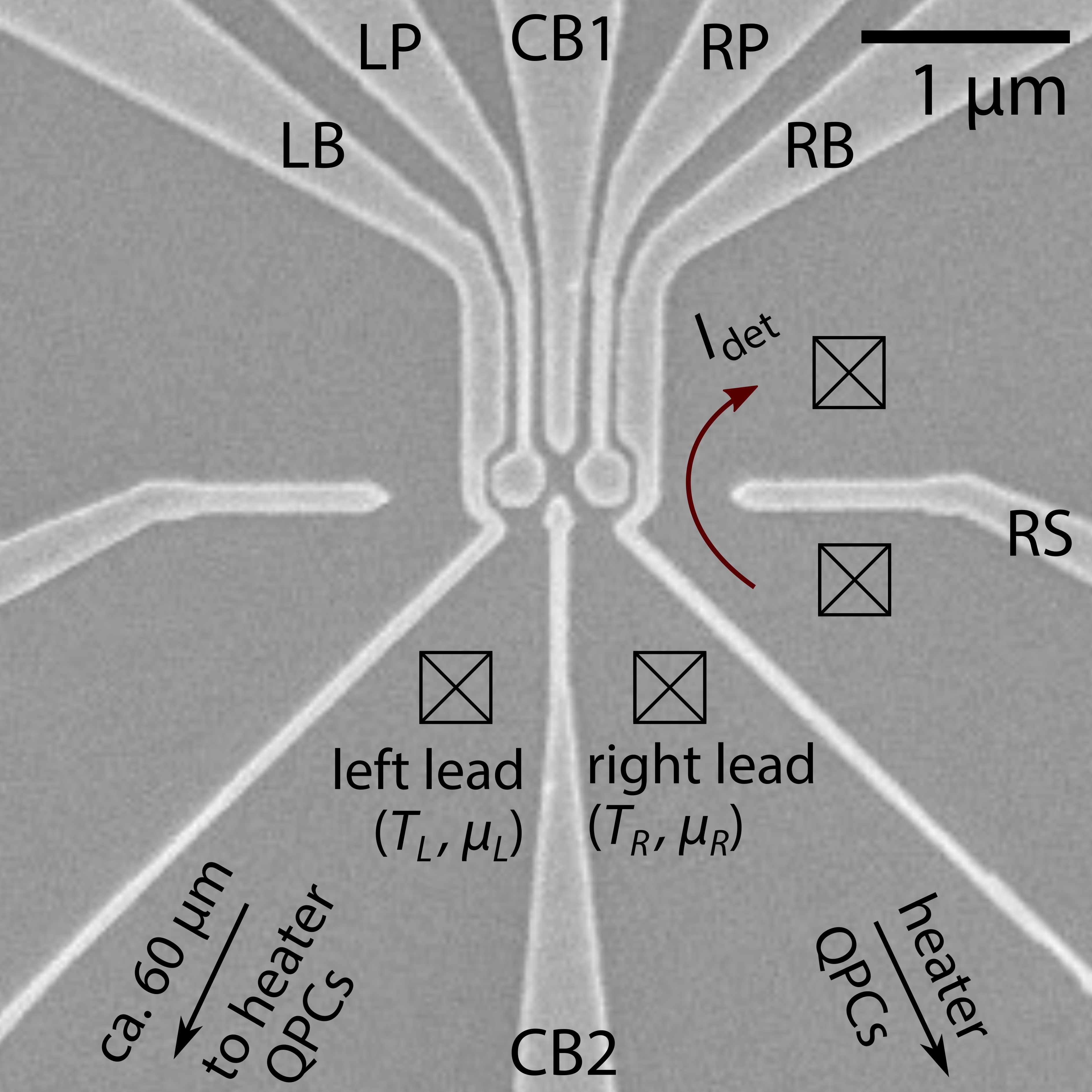}
    \caption{\label{fig:s1} Scanning electron micrograph of a nominally identical device (compare Fig.\@~1(a) of the main text) with all the gates used to form the dot and operate the device labeled.
    Not shown in this micrograph, but discussed in the text, is the split of each lead into two regions
    }
\end{figure}
\end{center}

The double quantum dot (DQD) is defined by six gates: the two plunger gates, LP and RP; two outer barriers, LB and RB; and two central barriers, CB1 and CB2.
A scanning electron micrograph of a nominally identical device is shown in Fig.~\ref{fig:s1}. 
Once the 2DEG is depleted beneath the gates---for isolated gates this occurs around $-400$ mV relative to the 2DEG---the coupling to the leads is controlled principally by the combination of one central barrier gate, CB2, and the outer barriers, LB and RB; the occupation by the plunger gates, LP and RP; and the interdot tunnel coupling by the other central barrier gate, CB1.
Voltage applied to any single gate affects multiple DQD parameters.
For example, the plunger gates mainly tune the dots' chemical potentials, but they also tune the interdot tunnel coupling and their respective dot--lead coupling, due to a combination of capacitive cross-talk and a shift in the dots' positions.

Besides the six gates that define the DQD, an additional gate, RS, can pinch off the current outside the RB gate to act as a QPC charge detector for both dots.
In what we present here, the QPC detector is tuned so that the conductance is between pinchoff and the first plateau.
We measure the current through the QPC using a current--voltage converter; the dc current is measured with a digital multimeter, the ac current with a lock-in amplifier.

\subsection{Heating the reservoirs}

Heating of the left and right reservoirs, aka the source and drain leads, essential to the measurement of the DQD's entropy, is accomplished in the following way.
Each reservoir is connected to two ohmic contacts.
These two ohmic contacts are in turn separated by two QPCs, used as constrictions, to increase the resistance between each reservoir's pair of contacts.
(When the DQD is defined, the reservoirs are isolated from each other by the center barrier gate CB2. The detector QPC reservoir is similarly isolated from the right reservoir by RB.)
Each reservoir can then be heated by the application of an ac current between the leads.
In these experiments, an ac voltage of a chosen amplitude, called the ``primary'' heater voltage, is applied at 40 Hz to one of the contacts to a reservoir, and another ac voltage, of a different amplitude, called the ``compensating'' heater voltage, is applied to the other contact on the same reservoir.
The compensating heater voltage is 180\textdegree~out of phase with the primary heater voltage, and its amplitude is chosen to minimize ac voltage fluctuation on the dot leads, as measured as ac current through the detector QPC by a lock-in amplifier.

\subsection{Minimizing voltage fluctuation due to the heating current}

\begin{center}
\begin{figure}[tb]
    \includegraphics[width= 0.9\textwidth]{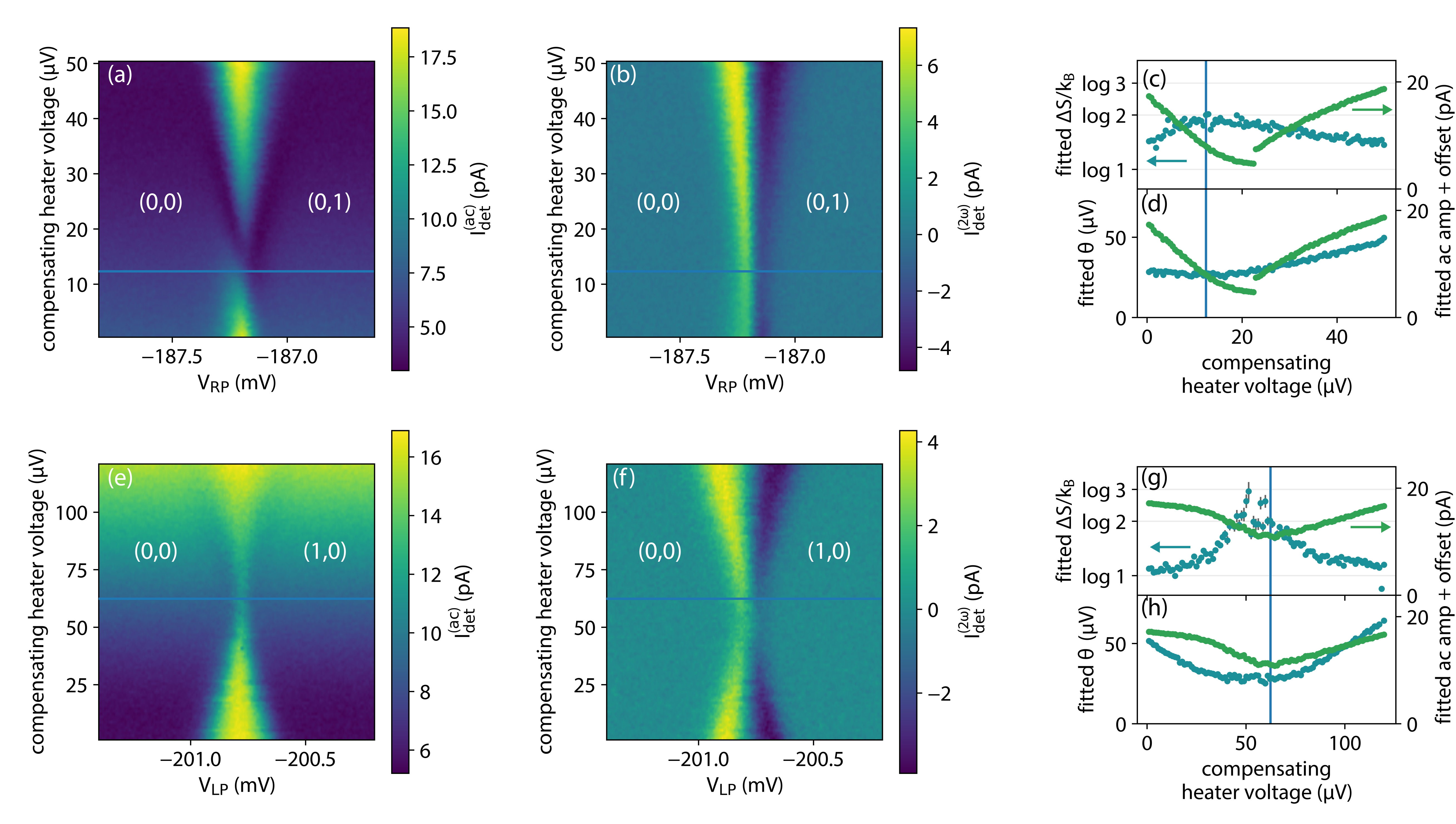}
    \caption{\label{fig:s2} Heater compensation.
    (a) $V_{\mathrm{RP}}$ sweeps across the $(0,0) \rightarrow (0,1)$ transition, heating on the right side, while the primary heater voltage is set at amplitude $100~\mu$V. (As always, the frequency of the applied voltage to the heater pair is 40 Hz.)
    The amplitude of the compensating heater voltage, which is 180\textdegree out of phase with respect to the primary heater voltage, is swept.
    We measure the current through the charge detector at the excitation frequency, $I_{\mathrm{det}}^{(\mathrm{ac})}$; here we plot the magnitude of the signal (lock-in R).
    (b) At the same time, we measure the current through the charge detector at the second harmonic of the heating excitation, $I_{\mathrm{det}}^{(2\omega)}$.
    From these data we can fit Eq.\@~1 of the main text to find (c) $\Delta S/k_\mathrm{B}$ and (d) $\theta$.
    Distortion of $I_{\mathrm{det}}^{(2\omega)}$ due to the direct driving of the current by the voltage fluctuations on the dot do not prevent fits, but they introduce error to the fitted values.
    The region in which the fitted values match our expectations matches a qualitative dip in the magnitude of the $I_{\mathrm{det}}^{(\mathrm{ac})}$ data, which we visualize on the righthand axes of (c) and (d) by plotting $I_0^{(\mathrm{ac})}+I^{(\mathrm{ac})}_{\mathrm{offset}}$ (see Eq.\@~\ref{eq:s_acfit}).
    (e-h) The same as (a-d) but for the $(0,0) \rightarrow (1,0)$ transition, using $V_\mathrm{LP}$ and heating on the left side. In this case the primary heater voltage is set at $120~\mu$V.
    }
\end{figure}
\end{center}

Measurements to calibrate the compensating voltage are shown in Fig.~\ref{fig:s2}.
We use a plunger gate to traverse a charge transition in the dot.
In Fig.~\ref{fig:s2}(a) the ac current through the detector at the heater modulation frequency, $I_{\mathrm{det}}^{(\mathrm{ac})}$, is plotted for different choices of the compensating heater voltage, while $V_{\mathrm{RP}}$ is used to traverse the $(N_\mathrm{L}, N_{\mathrm{R}}) = (0,0) \rightarrow (0,1)$ transition.
The magnitude of the primary heater voltage is fixed. 
In Fig.~\ref{fig:s2}(a), it is $100~\mathrm{\mu}$V.
The entropy can be measured, as discussed in the main text, from the second harmonic (of the ac heating voltage) of the current flowing through the detector QPC, $I_{\mathrm{det}}^{(2\omega)}$, which is plotted in Fig.~\ref{fig:s2}(b) for the same sweeps.
Our procedure is to choose a value for the compensating voltage that minimizes the size of the fluctuations in the detector current at the modulation frequency, i.e.\@ $I_{\mathrm{det}}^{(\mathrm{ac})}$.
We choose the value shown with a horizontal blue line in Figs.~\ref{fig:s2}(a) and (b).

How sensitive are our results to the choice of the compensating heater voltage?
Consider the behavior of the fitted parameters $\Delta S$ and $\theta$ as a function of the compensating heater voltage, shown in Figs.\ref{fig:s2}(c) and (d).
We see that $\Delta S$ forms a comparatively long, round maximum at the expected value, $k_\mathrm{B} \log 2$.
The behavior of $\theta$ is similar: the region where $\Delta S$ is maximal corresponds to a long, round minimum of $\theta$.
Neither depends sensitively on the choice of compensating voltage; ours is shown with a vertical blue line.
In Figs.\ref{fig:s2}(c) and (d), we also visualize the dip in size of $I_{\mathrm{det}}^{(\mathrm{ac})}$ as follows.
We observe in Fig.~\ref{fig:s2}(a), $I_{\mathrm{det}}^{(\mathrm{ac})}$ appears to have a peak at the transition.
We fit a gaussian to the absolute value of the lock-in R component of $I_{\mathrm{det}}^{(\mathrm{ac})}$, parameterized as
\begin{align}
|I_{\mathrm{det}}^{(\mathrm{ac})}| = I_0^{(\mathrm{ac})} \exp \left( -\left(\frac{V_{\mathrm{RP}}-V_{\mathrm{RP},0}}{\sigma}\right)^2\right) + I^{(\mathrm{ac})}_{\mathrm{offset}}.
\label{eq:s_acfit}
\end{align}
In Figs.~\ref{fig:s2}(c) and (d), we plot $I_0^{(\mathrm{ac})}+I^{(\mathrm{ac})}_{\mathrm{offset}}$.
We see that this quantity makes a sharper extremum than $\Delta S$ and $\theta$; we use it to check our choice of compensating voltage.

In Figs.~\ref{fig:s2}(e-h), we repeat the process for the left heater (and a transition on the left dot).
Since the resistance of the two constrictions on the left side is different, we find a different ratio of compensating voltages.

In principle it is possible to do finite-bias dc transport measurements in this setup by raising the dc bias of each pair of contacts relative to each other.
For the entropy measurements reported here, however, the dc bias difference between the pair of contacts to the left reservoir and the pair of contacts to the right reservoir is nominally zero.

\subsection{Averaging multiple traces}
In Fig.\@~1 of the main text, some of the traces have been averaged before fitting.
To do that, we fit each individual trace according to Eq.\@~(1) of the main text, shift each trace by the fitted value of $V_{\mathrm{mid}}$, interpolate each trace on a new (finer) grid, and average the resulting traces.
We then use these averaged data for fitting and integrating, as shown in Figs.\@~1(f-h) in the main text.
A characteristic improvement in signal-to-noise can be judged from the single (i.e.\@ no averaging) traces of Figs.\@~1(i-k) in the main text.
(The traces in Figs.\@~2 and 3 of the main text are not averaged.)

\subsection{Varying the dot--lead tunneling rates}

\begin{center}
\begin{figure}[tb]
    \includegraphics[width= 0.9\textwidth]{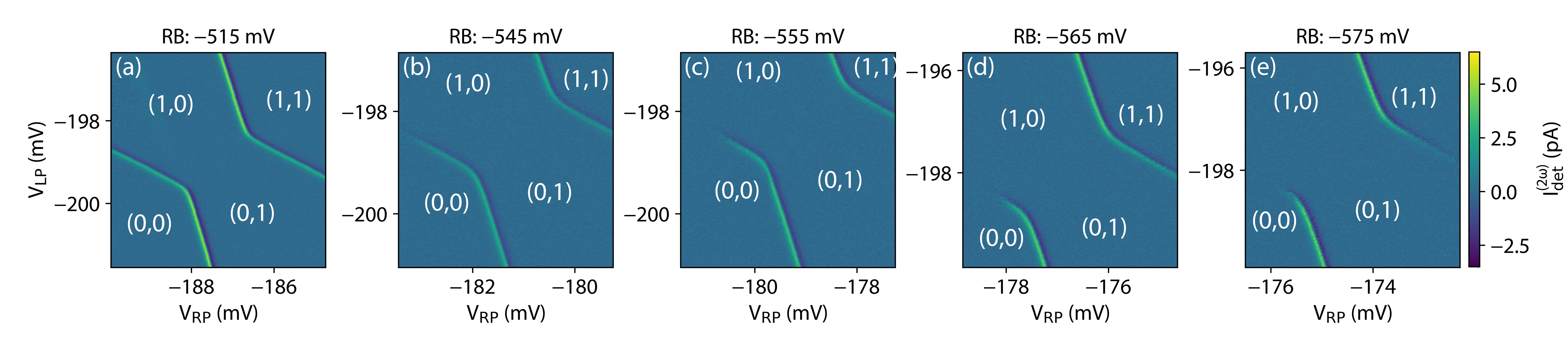}
    \caption{\label{fig:s3} As the tunneling rate to the right lead is reduced, the second harmonic current $I_{\mathrm{det}}^{(2\omega)}$ at the $N_\mathrm{L} = 0 \rightarrow 1$ transition fades. $I_{\mathrm{det}}^{(2\omega)}$ is plotted against $V_\mathrm{RP}$ and $V_\mathrm{LP}$ while $V_\mathrm{RB}$ is varied: (a) $V_\mathrm{RB} = -515~\mathrm{mV}$, 
    (b) $V_\mathrm{RB} = -545~\mathrm{mV}$,
    (c) $V_\mathrm{RB} = -555~\mathrm{mV}$,
    (d) $V_\mathrm{RB} = -565~\mathrm{mV}$, and
    (e) $V_\mathrm{RB} = -575~\mathrm{mV}$.
    }
\end{figure}
\end{center}

The tunneling rates between the DQD and the leads, $\Gamma_\mathrm{L}$ and $\Gamma_\mathrm{R}$ [see the schematic of Fig.~1(b)], are key parameters in the experiment, though we do not study their dependence in detail.
The important point is that Eq.\@~(1) of the main text holds in the thermally broadened one-lead regime, where the only relevant tunnel rate, whether $\Gamma_\mathrm{L}$ or $\Gamma_\mathrm{R}$, is much less than $k_\mathrm{B} T$.
As mentioned in the main text, additional coupling (i.e.\@ appreciable tunneling rate) to a second lead can be seen from  $I_{\mathrm{det}}^{(2\omega)}$.
In the case when the additional lead is cold, the signal simply fades away.
Figure~\ref{fig:s3} shows a sequence of measurements of $I_{\mathrm{det}}^{(2\omega)}$ at the same set of charge transitions with the right barrier voltage, $V_\mathrm{RB}$, tuned from $-515$ mV to $-575$ mV.
Pinching off this barrier has the primary effect of suppressing the tunneling rate to the right lead.

In Fig.~\ref{fig:s3}, the left lead is cold.
Since its temperature is unmodulated, we expect a suppression of any second harmonic signal due to DQD--left lead tunneling.
Across the range of plunger voltages here, the second harmonic begins to fade at the most negative values of $V_\mathrm{RP}$: $V_\mathrm{RP}$ also tunes the DQD--right lead tunnel rate, and this is where it is most suppressed.
As $V_\mathrm{RB}$ suppresses the DQD--right lead tunnel rate further, the point at which the second harmonic fades falls at smaller and smaller detuning---more and more positive values of $V_\mathrm{RP}$, in this case---and eventually, for $V_\mathrm{RB} = -575$ mV, $I_{\mathrm{det}}^{(2\omega)}$ is suppressed at nearly zero detuning.

\begin{center}
\begin{figure}[tb]
    \includegraphics[width= 0.9\textwidth]{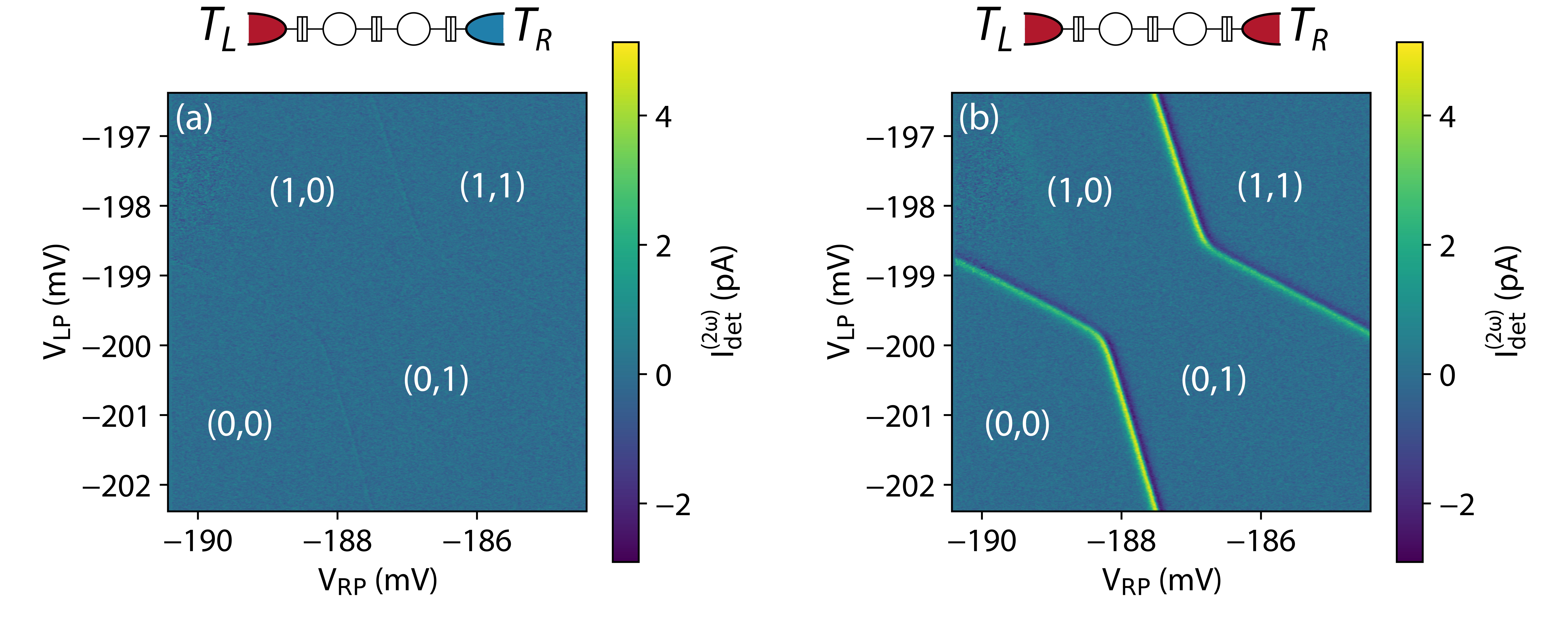}
    \caption{\label{fig:s4} When the coupling to one lead is weak enough, the temperature of that lead has very little or no effect on the second harmonic signal. 
    (a) The same experiment as Fig.~\ref{fig:s3}(a), but with the left lead hot. There is no visible signal.
    (b) The same experiment as Fig.~\ref{fig:s3}(a) and Fig.~\ref{fig:s4}(a), but with both leads heated. It matters only that the right lead is heated.
    }
\end{figure}
\end{center}

At the other end of the range, with $V_\mathrm{RB} = -515$ mV, the DQD--right lead tunneling rate dominates.
What happens if we modulate the left lead temperature under these conditions?
The answer is nothing that we can measure.
Figure~\ref{fig:s4} shows $I_{\mathrm{det}}^{(2\omega)}$ in that configuration with the left lead and both leads heated, respectively.
Since the DQD--left lead tunnel rate is much lower than the DQD--right lead tunnel rate, we see no effect of the left lead's heating.

\section{Single dot lever arm}
We measure the DQD under finite bias to find the lever arm $\alpha_{\mathrm{RR}}$, i.e.\@ for the right plunger acting on the right dot.
We apply a bias difference of size $|V_{\mathrm{SD}}| = 300~\mu$V between the left and right reservoirs. (For this experiment it is simplest to unplug the ``extra'' contacts that are needed for the heating; these pads are left floating.)
Then, from the dimensions of the resulting ``bias triangles'' we can determine the lever arms:
\begin{align}
\alpha_{\mathrm{LL}} = |V_{SD}/\delta V_{\mathrm{LP}}| \quad \& \quad \alpha_{\mathrm{RR}} = |V_{SD}/
\delta V_{\mathrm{RP}}|.
\end{align}
In this way we find $\alpha_{\mathrm{RR}} = 0.13$ and $\alpha_{\mathrm{LL}} = 0.11$.
We estimate an error of 10\% due to the identification of the triangles' vertices and drift of the dot parameters as the DQD is tuned to into different configurations.
This lever arm is used to determine all the absolute temperatures in Figs.\@~1 and 2 of the main text, along with the associated discussion.
The demonstration that the DQD remains at constant temperature as the plunger gates are tuned past the triple point, however [Figs.\@~2(e) and 3(d) of the main text], rely only the lever arm ratio, which we determine from the charge stability diagram directly.

\begin{center}
\begin{figure}[tb]
    \includegraphics[width=0.9\textwidth]{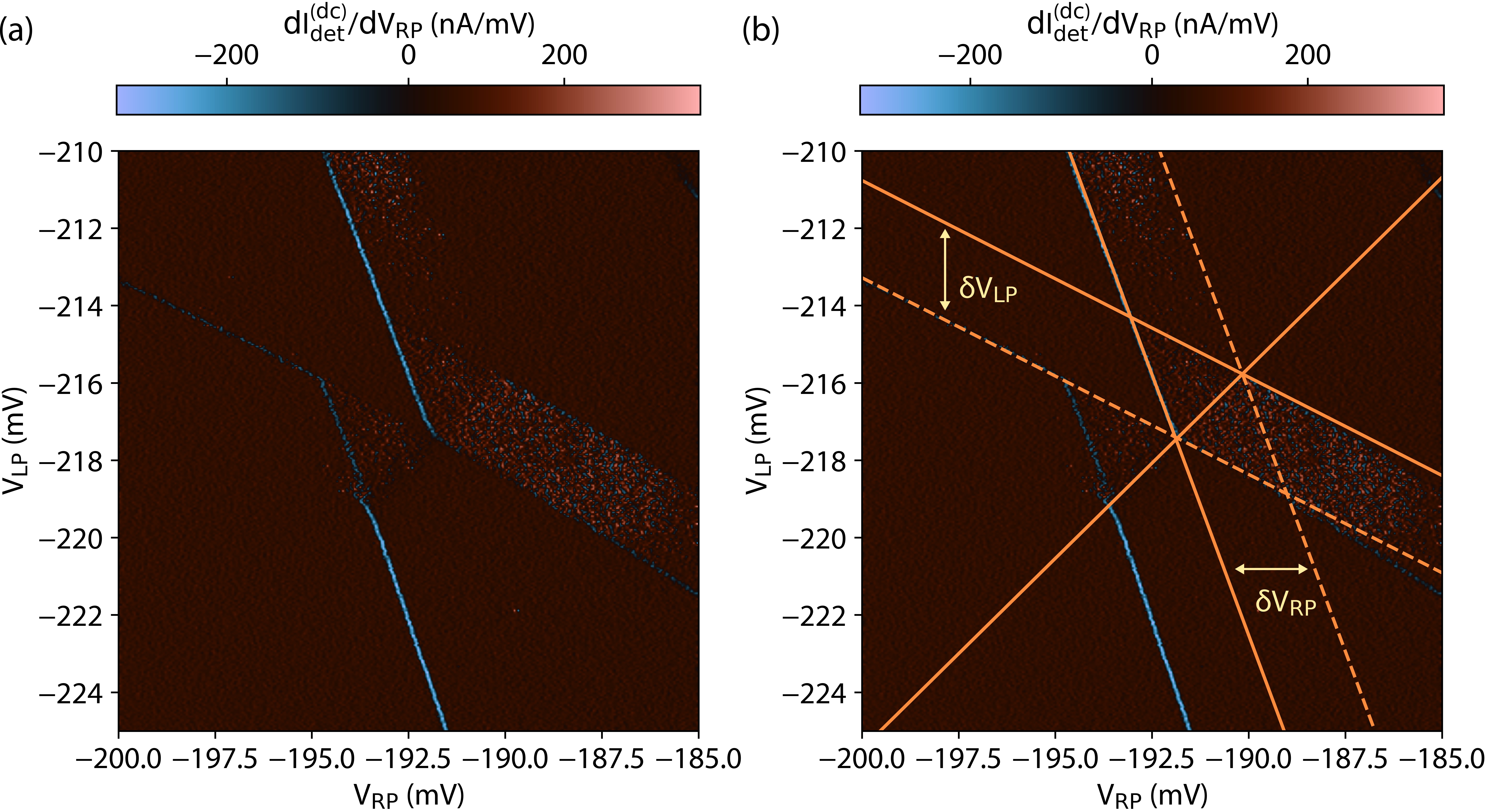}
    \caption{\label{fig:s5} Finite bias measurement, with $|V_{\mathrm{SD}}| = 300~\mu$V, as seen in the dc current through the charge detector. From the dimensions of the triangles we determine $\delta V_{\mathrm{LP}}$ and $\delta V_{\mathrm{RP}}$, from which we can determine $\alpha_{\mathrm{LL}}$ and $\alpha_{\mathrm{RR}}$.
    (a) Detector dc current, differentiated with respect to the right plunger voltage, $dI_{\mathrm{det}}^{\mathrm{(dc)}}/dV_{\mathrm{RP}}$, plotted against $V_{\mathrm{RP}}$ and $V_{\mathrm{LP}}$.
    (b) The same as (a), with overlay demonstrating how $\delta V_{\mathrm{LP}}$ and $\delta V_{\mathrm{RP}}$ are determined.    
    }
\end{figure}
\end{center}

\section{Extended data for Figure 2(f)and 2(g)}

\begin{center}
\begin{figure}[tb]
    \includegraphics[width=0.9\textwidth]{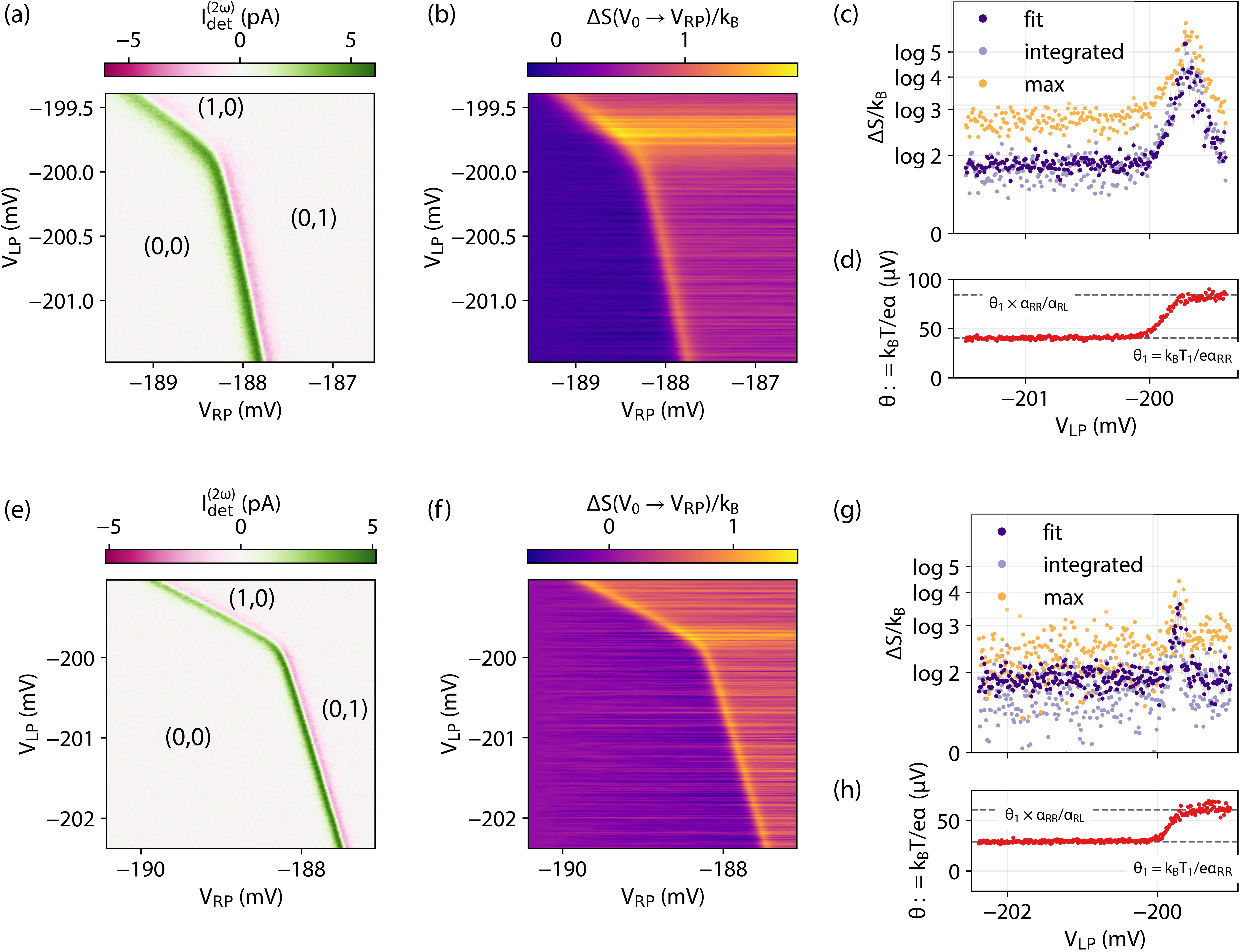}
    \caption{\label{fig:s7} Extended data for Fig.~2(f-g) of the main text. The data shown in Fig.~2 of the main text have $T_1 = 48 $ mK.
    Near the triple point, first carrier transition, for $T_1 = 62$ mK,
    (a) $I^{(2\omega)}_{\mathrm{det}}$ is plotted as a function of $V_{\mathrm{RP}}$ and $V_{\mathrm{LP}}$.
    (b) Cumulative integral of the entropy with respect to $V_{\mathrm{RP}}$, starting from $(N_\mathrm{L}, N_\mathrm{R}) = (0,0)$.
    At every value of $V_{\mathrm{LP}}$, the product $I_0 \theta$ is found from a fit to $I^{(2\omega)}_{\mathrm{det}}(V_{\mathrm{RP}})$.
    (c) Fitted entropy change across the transition for every $V_{\mathrm{LP}}$ in (a).
    The peak value for each integration trace is also plotted, which records an extra degeneracy of one throughout, as well as the value found from integration.
    (d) Fitted $\theta = k_\mathrm{B} T/ e \alpha$ as a function of $V_{\mathrm{LP}}$.
    Also shown is a dotted line showing an average value $\theta_1$ of $\theta$ at the $(0,0)\rightarrow (0,1)$ transition, far from the triple point, and a dotted line showing the calculated value $\theta_1 \times \alpha_{\mathrm{RR}}/\alpha_{RL}$.
    The ratio of lever arms is found independently, using the charge stability diagram.
    (e-h) Same as (a-d), with $T_1 = 45$ mK.
    }
\end{figure}
\end{center}

In the main text, Figs.\@~2(f) and (g) refer to two data sets, not shown in Figs.\@~2(a-e).
In Fig.~\ref{fig:s7} we provide, for completeness, the data and analysis from which the peak widths and heights were extracted and presented in Figs.\@~2(f) and (g).
The interpretation of these data is the same as that discussed in the main text.

\section{Extended data for Pauli blockade configuration}
\begin{center}
\begin{figure}[tb]
    \includegraphics[width=0.9\textwidth]{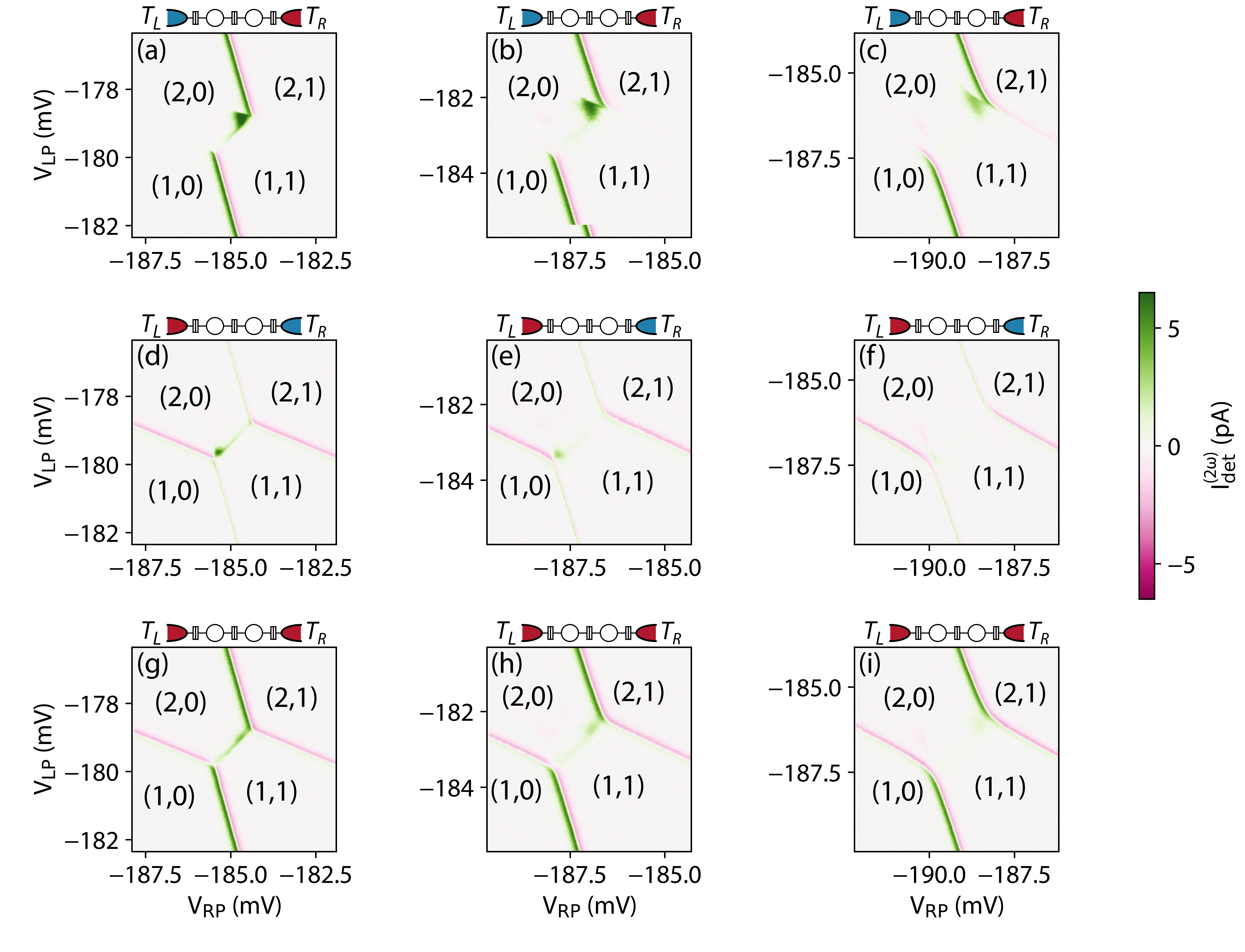}
    \caption{\label{fig:s6} Triangles switch sides with temperature gradient.
    (a) The right reservoir is hot, as depicted schematically; we plot $I_{\mathrm{det}}^{(2\omega)}$.
    (b-c) The same as (a), with progressively increasing $V_{\mathrm{CB1}}$ (see Fig.~\ref{fig:s1}) to increase the interdot tunnel rate $\Gamma_{\mathrm{int}}$ (see Fig.~1 of the main text).
    (d-f) The same experiment as (a-c) with the left reservoir hot. In this configuration, there is nevertheless still a small heating current applied to the right reservoir, which is why the right dot transitions are visible.
    (g-i) The same as (a-c) with both reservoirs heated.
    }
\end{figure}
\end{center}

The triangular features shown in Fig.\@~3 of the main text and discussed appear for a wide range of interdot tunnel couplings.
Only one is shown in the main text.
Figure~\ref{fig:s6} shows data for three different interdot tunnel rates, with one reservoir hot, the other reservoir hot, and both reservoirs hot.
As discussed in the main text, the switch of heated reservoir causes the triangle to switch sides---in other words, it appears near the $(1,0),(1,1),(2,0)$ triple point when the right reservoir is heated, but near the $(2,0),(1,1),(2,1)$ triple point when the left reservoir is heated.
When both reservoirs are heated, the triangles disappear.

\section{Rate equation model}
In this section we introduce the rate equation model we use for the calculations in Fig.\@~3 of the main text, as well as to understand the importance of specific DQD parameters to the presence of the triangular features seen there.

In our experiment, we use the current through a QPC to detect the charge on our DQD, which is tunnel-coupled to two different leads.
This is the system we seek to understand with the help of the rate equation model.
We write the ``detector'' current $I_{\mathrm{det}}^{(\mathrm{dc})}$ through the QPC detector as a linear combination of the average occupation $p_{\mathrm{L/R}}$ of each dot:
\begin{align}
I_{\mathrm{det}}^{(\mathrm{dc})} = I_{\mathrm{L}}\,p_{\mathrm{L}}+I_{\mathrm{R}} \,p_{\mathrm{R}}, \label{eq:s_model_current}
\end{align}
where $I_{\mathrm{L}(\mathrm{R})}$ quantifies the detector sensitivity with respect to changes in occupation on the left (right) dot (compare Fig.\@~1(b) of the main text).

The probabilities of being in the left or right dot, $p_\mathrm{L}$ or $p_\mathrm{R}$, we separate according to state.
Our model has three states, labeled $0,~1,$ and $2$.
State $0$ is an empty state (note that it results in zero current through the detector); states $1$ and $2$ are single-particle states.
For the system Hamiltonian we take that of a single electron in a DQD, with onsite energy $\epsilon$, detuning $\delta$, and tunnel coupling $t$. In the left--right basis, we write the Hamiltonian as
\begin{align}
H &= \begin{pmatrix} \epsilon + \delta/2 & t \\ t & \epsilon - \delta/2\end{pmatrix}; \label{eq:s_ham}
\end{align}
the ground (bonding) state of this system is state $1$, the excited (antibonding) state of this system is state $2$, and the empty state, with energy $0$, is state $0$.
The resulting energy eigenvalues are
\begin{align}
    E_{1,2} = \epsilon \pm \frac{1}{2}\sqrt{\delta^2 + 4t^2};
\end{align}
the corresponding normalized eigenvectors are (in the left--right basis):
\begin{subequations}
\begin{align}
\ket{1} &= \begin{pmatrix} \cos(\theta/2) \\ \sin(\theta/2) \end{pmatrix} \\
\ket{2} &= \begin{pmatrix} -\sin(\theta/2) \\ \cos(\theta/2)\end{pmatrix},
\end{align}
\end{subequations}
where $\cot(\theta)=\delta/2t$ defines the mixing angle $\theta\in[0,\pi]$.
Thus we can rewrite Eq.~(\ref{eq:s_model_current}) in terms of the probabilities of being in each of the three states, $p_0$, $p_1$, and $p_2$:
\begin{align}
\label{eq:s_curr2}
    I_{\mathrm{det}}^{(\mathrm{dc})} = I_{\mathrm{L}}\, \left(p_1 \,\cos^2(\theta/2) + p_2 \,\sin^2(\theta/2)\right) + I_{\mathrm{R}} \, \left(p_1 \,\sin^2(\theta/2) + p_2 \,\cos^2(\theta/2)\right).
\end{align}

The probabilities $p_0$, $p_1$, and $p_2$ obey the rate equation (and, separately, must sum to $1$).
We define the rate $W_{ij}$ as the rate for a transition from state $j$ to state $i$.
The stationary probabilities solve this system of equations:
\begin{subequations}
    \begin{align}
p_0 + p_1 + p_2 &= 1 \\
W_{10} \, p_0 - (W_{01}+W_{21})\, p_1 + W_{12} \, p_2 &=0 \\
W_{20} \, p_0 + W_{21} \, p_1 - (W_{02} + W_{12})\, p_2 &= 0.
    \end{align}
\end{subequations}
Explicitly, in terms of the rates, then:
\begin{subequations}
\label{eq:s_probs}
    \begin{align}
        p_0 &= \frac{W_{01} W_{02} + W_{01} W_{12} + W_{02} W_{21}}{W^2} \\
        p_1 &= \frac{W_{02} W_{10} + W_{12} W_{10} + W_{12} W_{20}}{W^2} \\
        p_2 &= \frac{W_{01} W_{20} + W_{10} W_{21} + W_{20} W_{21}}{W^2},
    \end{align}
\end{subequations}
with
\begin{align}
    W^2 &= W_{01}(W_{02} + W_{12} + W_{20}) + W_{02}(W_{10} + W_{21}) + (W_{10} + W_{20})(W_{12} + W_{21}).
\end{align}

The rates themselves we can write more transparently, i.e.\@ in terms of the bare dot--lead tunnel rates, $\Gamma_\mathrm{L}$ and $\Gamma_\mathrm{R}$, and the  lead temperatures, $T_\mathrm{L}$ and $T_\mathrm{R}$, which give rise to the following Fermi-Dirac distributions in the leads:
\begin{align}
    f_{\mathrm{L}/\mathrm{R}}(E) = \frac{1}{e^{E/k_\mathrm{B} T_{\mathrm{L}/\mathrm{R}}} + 1}.
\end{align}
The rates are then
\begin{subequations}
\label{eq:s_rates}
    \begin{align}
        W_{10} &= d_1 \left(\Gamma_\mathrm{L} f_{\mathrm{L}}(E_1) \cos^2(\theta/2) + \Gamma_{\mathrm{R}} f_\mathrm{R} (E_1)\sin^2(\theta/2) \right) \\
        W_{01} &= d_0 \left( \Gamma_\mathrm{L} (1- f_{\mathrm{L}}(E_1)) \cos^2(\theta/2) + \Gamma_{\mathrm{R}} (1-f_\mathrm{R} (E_1))\sin^2(\theta/2)\right) \\
        W_{20} &= d_2 \left( \Gamma_\mathrm{L} f_{\mathrm{L}}(E_2) \sin^2(\theta/2) + \Gamma_{\mathrm{R}} f_\mathrm{R} (E_2)\cos^2(\theta/2)\right) \\
        W_{02} &= d_0 \left( \Gamma_\mathrm{L} (1- f_{\mathrm{L}}(E_2)) \sin^2(\theta/2) + \Gamma_{\mathrm{R}} (1-f_\mathrm{R} (E_2))\cos^2(\theta/2)\right) \\
        W_{21} &= 0 \\
        W_{12} &= \Gamma_{\mathrm{r}},
    \end{align}
\end{subequations}

where we have added $\Gamma_\mathrm{r}$, the relaxation rate from the antibonding $2$ state to the bonding $1$ state, by hand.
We have also introduced notation to describe the degeneracy of each state $d_i$.
With the exception of the relaxation rate, whose effect we will discuss soon, the other expressions all say the same thing: the rate at which an electron enters the occupied DQD states ($1$ and $2$) is determined by the rate of tunneling from the leads into the left or right dot weighted by the probability of the electronic eigenstate in the respective dot.
To change those rates, we change the bare rates $\Gamma_\mathrm L$ or $\Gamma_\mathrm R$ (in the experiment, how pinched off the leads are with respect to the dot), the temperature of the leads, or the energetics of the dot ($t$ or $\delta$), which affect both the Fermi-Dirac terms as well as the projections like $\cos^2(\theta/2)$.
(We treat the chemical potentials of the leads as fixed, as in the experiments.)

Finally, the experiment measures the second harmonic of the heating current, $I_{\mathrm{det}}^{(2\omega)}$.
Since this quantity is proportional to $\partial I_{\mathrm{det}}^{(\mathrm{dc})} / \partial T\approx \Delta I_{\mathrm{det}}^{(\mathrm{dc})} / \Delta T$, we calculate the quantity
\begin{align}
    \Delta I_{\mathrm{det}}^{(\mathrm{dc})} &= I_{\mathrm{det}}^{(\mathrm{dc})} (T_{\mathrm{mod}} + T_{\mathrm{L}/\mathrm{R}}) - I_{\mathrm{det}}^{(\mathrm{dc})}( T_{\mathrm{L}/\mathrm{R}}) \\ &\propto I_{\mathrm{det}}^{(2\omega)},
\end{align}
where $T_\mathrm{mod}$ is the modulated temperature of the lead over the temperature of the lead, which is itself $T_\mathrm{mod}$ over the base (electron) temperature of the fridge $T_\mathrm{base}$.
The calculation runs by evaluating Eq.\@~(\ref{eq:s_curr2}) with Eq.\@~(\ref{eq:s_rates}) and Eq.\@~(\ref{eq:s_probs}).

\subsection{Results and discussion}
\begin{center}
\begin{figure}[tb]
    \includegraphics[scale= 1]{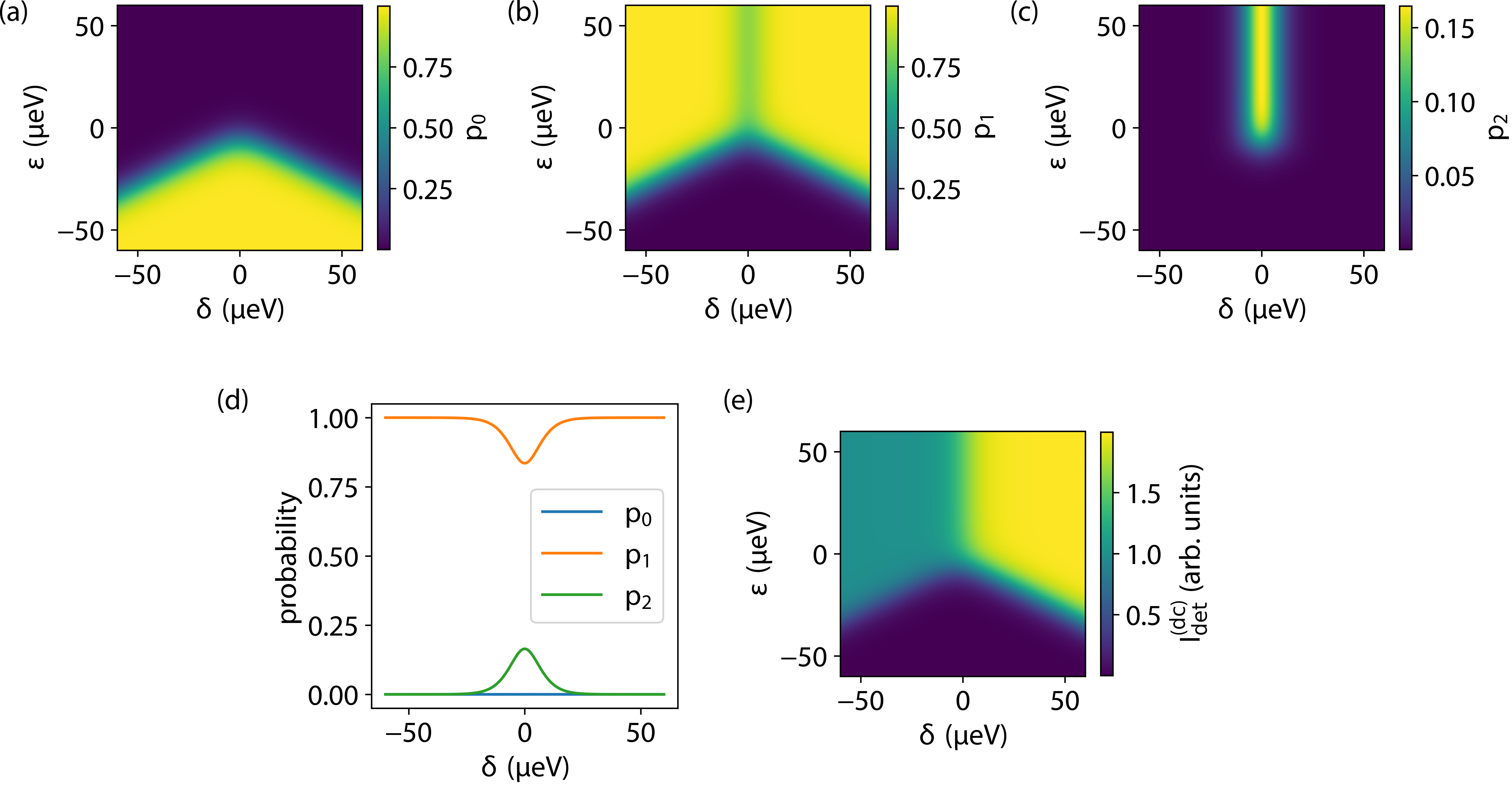}
    \caption{\label{fig:s8_1} Probabilities and calculated dc current. 
    Calculated (a) $p_0$, (b) $p_1$, and (c) $p_2$ plotted against $\epsilon$ and $\delta$ for $\Gamma_\mathrm{L} = \Gamma_{\mathrm{R}} = 1$ mK. The degeneracies are $d_0 = 1, d_1 = d_2 =2$.
    The temperatures are $T_\mathrm{L} = T_{\mathrm{R}} = 50$ mK, and the interdot tunnel coupling $t=7$ mK.
    (d) Linecut of the data in (b) and (c) at $\epsilon = 30~\mu$eV.
    (e) Simulated $I_\mathrm{det}^{(\mathrm{dc})}$, with $I_\mathrm{L} = 1$ and $I_\mathrm{R} = 2$.
    }
\end{figure}
\end{center}

A basic demonstration of the model at dc is shown in Fig.~\ref{fig:s8_1}.
The first three panels, Figs.~\ref{fig:s8_1}(a-c), show $p_0,$ $p_1,$ and $p_2$ plotted against $\epsilon$ and $\delta$ [see Eq.~(\ref{eq:s_ham})].
For sufficiently low $\epsilon$, the empty $0$ state is occupied, as shown in Fig.~\ref{fig:s8_1}(a).
At higher $\epsilon$, state $1$ or $2$ is occupied.
As can be seen from Fig.~\ref{fig:s8_1}(b) and (c), there is a narrow strip near $\delta = 0$ where there is nonzero probability of occupying the excited state $2$; otherwise, state $1$ is occupied.
For another view of the same thing, Fig.~\ref{fig:s8_1}(d) shows a line cut across the data shown in Figs.~\ref{fig:s8_1}(b) and (c) at $\epsilon = 30~\mu$eV.
Finally, for a view of the same thing in the left--right basis, Fig.~\ref{fig:s8_1}(e) shows a plot of the calculated detector current $I_\mathrm{det}^{(\mathrm{dc})}$, with $I_\mathrm{L} = 1$ and $I_\mathrm{R} = 2$ [see Eq.~(\ref{eq:s_model_current})].
(Elsewhere we choose $I_\mathrm{R}/I_\mathrm{L} = 5/3$ to be similar to our experimental values. Compare Fig.~1(f) and 1(i) in the main text.)

\begin{center}
\begin{figure}[tb]
    \includegraphics[scale= 1]{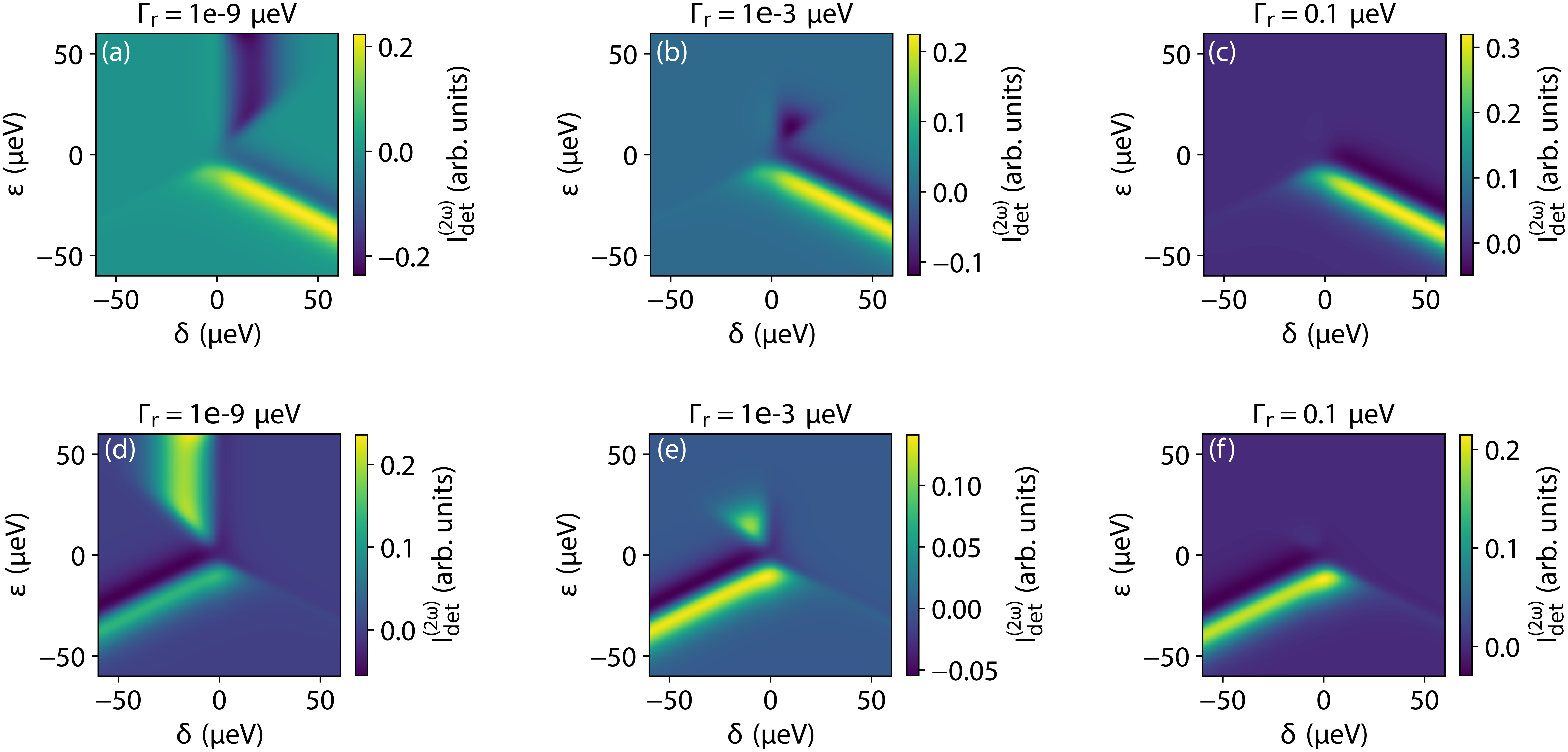}
    \caption{\label{fig:s8_2} Triangular features' dependence on relaxation rate. 
    $I_{\mathrm{det}}^{(2\omega)}$, i.e.~\@ $\Delta I_{\mathrm{det}}^{(\mathrm{dc})}$, is plotted against $\epsilon$ and $\delta$ for varying $\Gamma_\mathrm{r}$, with (a-c) 20 mK modulation added to $T_\mathrm{R}$ and 0.2 mK modulation added to $T_\mathrm{L}$ and (d-f) 20 mK modulation added to $T_\mathrm{L}$ and 0.2 mK modulation added to $T_\mathrm{R}$.
    The other calculation parameters are $d_0 = 1$, $d_1 = d_2 = 2$, $I_\mathrm{L} = 1$, $I_\mathrm{R} = 5/3$, $\Gamma_\mathrm{L} = \Gamma_\mathrm{R} = 1$ mK, $t = 7$ mK, and a base temperature of $T_\mathrm{L} = T_\mathrm{R} = 20$ mK.
    }
\end{figure}
\end{center}

In Fig.~\ref{fig:s8_2} is plotted the calculated second harmonic current, $I_{\mathrm{det}}^{(2\omega)}$, i.e.~\@ $\Delta I_{\mathrm{det}}^{(\mathrm{dc})}$. (In this section we use the two interchangeably.)
These data complement those shown in Fig.~3 of the main text.
The key point is that, with one side heated, the size of the triangular feature is controlled by the relaxation rate $\Gamma_\mathrm{r} = W_{12}$, which we add in by hand.
Figures~\ref{fig:s8_2}(a-c) show $I_{\mathrm{det}}^{(2\omega)}$ for different values of the relaxation rate with the right reservoir heated; Figs.~\ref{fig:s8_2}(d-f) show the same with the left reservoir heated.
One key thing is that the sign of the triangular feature changes as the heated reservoir is changed.

\begin{center}
\begin{figure}[tb]
    \includegraphics[scale= 1]{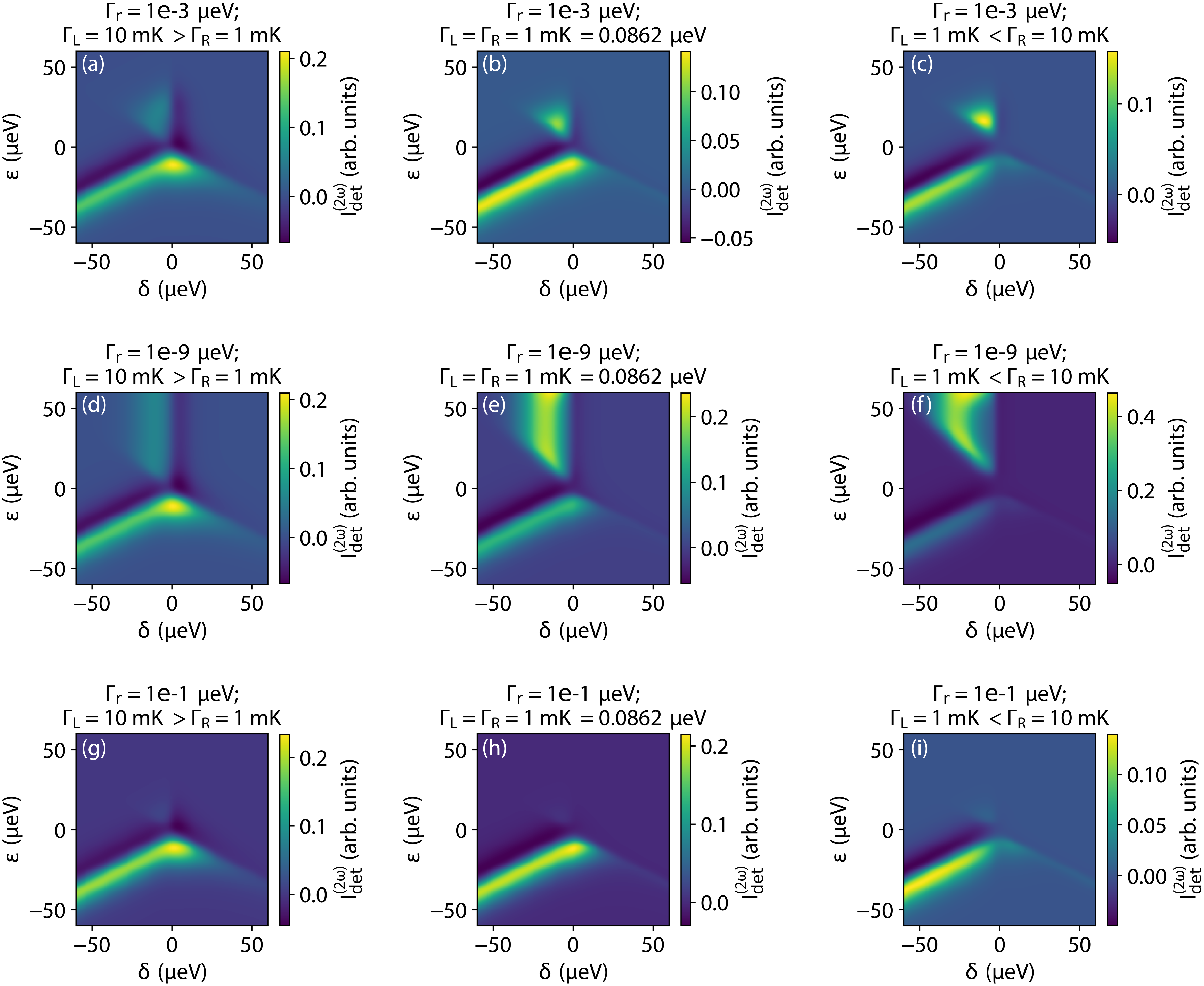}
    \caption{\label{fig:s8_3} Asymmetric dot--lead coupling is relatively unimportant to the triangles' visibility. 
    $I_{\mathrm{det}}^{(2\omega)}$, i.e.~\@ $\Delta I_{\mathrm{det}}^{(\mathrm{dc})}$, is plotted against $\epsilon$ and $\delta$ for varying $\Gamma_\mathrm{R}$ and $\Gamma_\mathrm{L}$, with 20 mK modulation added to $T_\mathrm{L}$ and 0.2 mK modulation added to $T_\mathrm{R}$, and (a-c) $\Gamma_\mathrm{r} =1$e-$3~\mu$eV,
    (d-f) $\Gamma_\mathrm{r} =1$e-$9~\mu$eV,
    and (g-i) $\Gamma_\mathrm{r} =1$e-$1~\mu$eV. 
    The other calculation parameters are $d_0 = 1$, $d_1 = d_2 = 2$, $I_\mathrm{L} = 1$, $I_\mathrm{R} = 5/3$, $t = 7$ mK, and a base temperature of $T_\mathrm{L} = T_\mathrm{R} = 20$ mK.
    }
\end{figure}
\end{center}

Given the asymmetry of the heating required for the triangles to be visible (see Fig.~3 of the main text), is it possible that the triangles are just as much a feature of asymmetric dot--lead coupling?
Figure~\ref{fig:s8_3}, in which the calculated $I_{\mathrm{det}}^{(2\omega)}$ is plotted for different values of $\Gamma_\mathrm{L,R}$ and $\Gamma_\mathrm{r}$, shows that the answer is no, essentially.
While approximately 10$\times$ changes in the size of $\Gamma_\mathrm{L}$ with respect to $\Gamma_\mathrm{R}$ affect the detailed appearance of the triangles, the presence of the triangles does not require and cannot be brought about by only unbalanced dot--lead tunnel coupling.

\begin{center}
\begin{figure}[tb]
    \includegraphics[scale= 1]{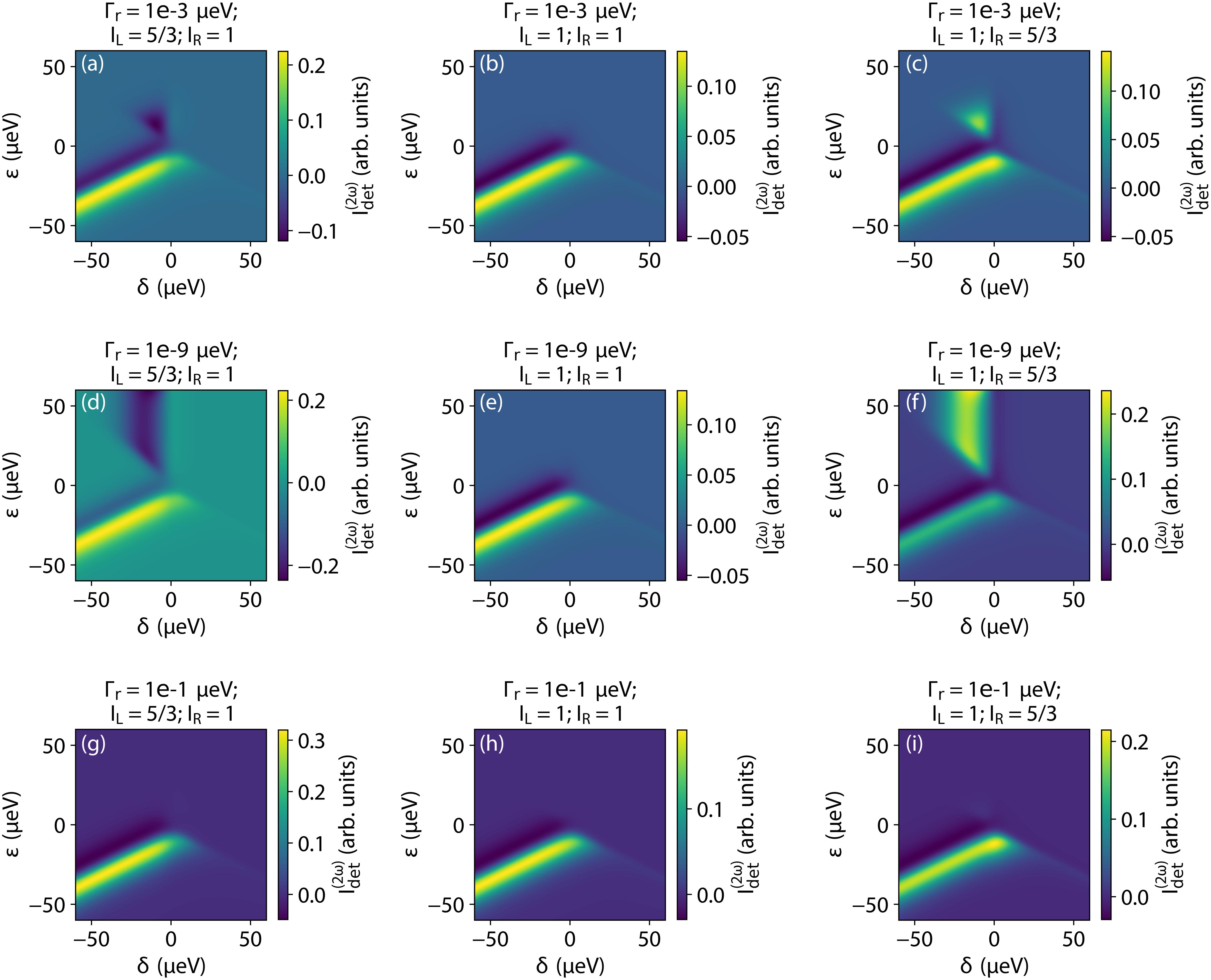}
    \caption{\label{fig:s8_4} Asymmetric detector sensitivity coupling is important to the triangles' visibility. 
    $I_{\mathrm{det}}^{(2\omega)}$, i.e.~\@ $\Delta I_{\mathrm{det}}^{(\mathrm{dc})}$, is plotted against $\epsilon$ and $\delta$ for varying $I_\mathrm{R}$ and $I_\mathrm{L}$, with 20 mK modulation added to $T_\mathrm{L}$ and 0.2 mK modulation added to $T_\mathrm{R}$, and (a-c) $\Gamma_\mathrm{r} =1$e-$3~\mu$eV,
    (d-f) $\Gamma_\mathrm{r} =1$e-$9~\mu$eV,
    and (g-i) $\Gamma_\mathrm{r} =1$e-$1~\mu$eV. 
    The other calculation parameters are $d_0 = 1$, $d_1 = d_2 = 2$, $\Gamma_\mathrm{L} = \Gamma_\mathrm{R} = 1$ mK, $t = 7$ mK, and a base temperature of $T_\mathrm{L} = T_\mathrm{R} = 20$ mK.
    }
\end{figure}
\end{center}

What can fully suppress (or create) the triangles, given sufficient relaxation rate, is the asymmetry of the detector response, which we model with $I_\mathrm{L}$ and $I_\mathrm{R}$.
As mentioned above, we use $I_\mathrm{R}/I_\mathrm{L} =5/3$ in Fig.~3 of the main text and (unless otherwise specified) elsewhere, since it approximately matches the experimental situation.
What if the experimental situation were reversed, or what if we had a perfectly fair detector?
In Fig.~\ref{fig:s8_4} we plot $I_{\mathrm{det}}^{(2\omega)}$ for different values of $I_\mathrm{L}$ and $I_\mathrm{R}$.
Even for configurations $\Gamma_\mathrm{r} = 1\text{e-}9, 1\text{e-}3$ eV, where we know the triangles are visible, a detector for which $I_\mathrm{L} = I_\mathrm{R}$ does not see any triangular features.

On the basis of Figs.~\ref{fig:s8_2}--\ref{fig:s8_4} we claim in the main text that the asymmetry of the detector response and a sufficiently low relaxation rate are required for the observation of these triangular features.
Given that the difference in sensitivity of the detector with respect to the charging of each dot is approximately fixed throughout our experiments, and given that the more frequent observation is the lack of such triangular features, we claim the triangles' presence in the data shown in Fig.~3 of the main text is due to a suppression of the relaxation rate.

\end{document}